\documentclass[11pt,leqno]{article}
\usepackage{graphicx}
\usepackage{amssymb}
\usepackage{amsmath}
\usepackage{pifont}
\usepackage{subfig}
\usepackage[dvipsnames,usenames]{color}

\usepackage{tabularx}
\usepackage{multirow}
\usepackage{multicol}
\usepackage{eepic,epic}
\usepackage{epsfig}
\usepackage{color}
\usepackage{mathptmx}
\usepackage{amsthm}
\usepackage{nicefrac}
\usepackage{booktabs}
\usepackage{todonotes}
\usepackage{tikz}
\usetikzlibrary{positioning}
\usetikzlibrary{calc}
\usepackage{enumerate}
\usepackage{rotating}
\usepackage[linesnumbered,vlined,ruled]{algorithm2e}
\input epsf %

\newcommand{\ML}{ML}

\newenvironment{proofs}{\noindent{\bf Proof.}\hspace*{1em}}{\literalqed\bigskip}
\newcommand\qedblob{\mbox{\ding{113}}}
\def\literalqed{{\ \nolinebreak\hfill\mbox{\qedblob\quad}}}

\newtheorem{proposition}{Proposition}

\begin{document}

\title{Randomized Classifiers vs Human Decision-Makers: Trustworthy AI May Have to Act Randomly and Society Seems to Accept This}

\author{ G\'{a}bor Erd\'{e}lyi \\
        University of Canterbury\\
       School of Mathematics and Statistics\\
       Christchurch, New Zealand\\
       Corvinus University of Budapest\\
       Budapest, Hungary
       \and
        Olivia J. Erd\'elyi  \\
        University of Canterbury\\
       School of Law\\ 
       Christchurch, New Zealand \\
            Soul Machines \\
            Auckland, New Zealand\\
            Corvinus University of Budapest\\
            Budapest, Hungary
       \and
        Vladimir Estivill-Castro \\
        Universitat Pompeu Fabra\\
       Departament de Tecnologies de la Informaci\'{o} i les Comunicacions \\
		Barcelona, Spain}

\date{November 15, 2021}
\maketitle

\begin{abstract}
As \emph{artificial intelligence} (AI) systems are increasingly involved in decisions affecting our lives, ensuring that automated decision-making is fair and ethical has become a top priority.  Intuitively, we feel that akin to human decisions,  judgments of artificial agents should necessarily be grounded in some moral principles.  Yet a decision-maker (whether human or artificial) can only make truly ethical (based on any ethical theory)  and fair (according to any notion of fairness) decisions if full information on all the relevant factors on which the decision is based are available at the time of decision-making. This raises two problems: (1) In settings, where we rely on AI systems that are using classifiers obtained with supervised learning, some induction/generalization is present and some relevant attributes may not be
present even during learning. (2) Modeling such decisions as games reveals that any---however ethical---pure strategy is inevitably susceptible to exploitation.
 Moreover, in many games, a Nash Equilibrium can only be obtained by using mixed strategies, i.e., to achieve mathematically optimal outcomes,  decisions must be randomized.  In this paper, we argue that in supervised learning settings, there exist random classifiers that perform at least as well as deterministic classifiers, and may hence be the optimal choice in many circumstances. We support our theoretical results with an empirical study indicating a positive societal attitude towards randomized artificial decision-makers, and discuss some policy and implementation issues related to the use of random classifiers that relate to and are relevant for current AI policy and standardization initiatives.
\end{abstract}

\section{Introduction}
\label{intro}

Among a society of equals, decisions  affecting members' lives are of the utmost importance. Accordingly, in the pre-AI age, it was a widely accepted---but unfortunately less widely practiced---principle  that society should entrust carefully selected human decision-makers with such choices.  Ideally, those positions would be associated with strict requirements on appropriate expertise and moral integrity,  as well as rigorous accountability mechanisms, to provide sufficient assurance for society to trust the judgment of those in charge.

Over time, however, technological innovation has enabled the automation of a growing number of tasks in, e.g.,  production processes, transport, and the information and communication sector. Successful attempts of automation radically improved human performance, significantly raising humanity's welfare and continuously changing our way of life. Most episodes of innovation have entailed more or less severe temporary disruptions until the economy adjusted to the effects of innovation,  society has become familiar with the new technologies in question and either grew to trust them or rejected their adoption due to negative experiences. 

Recent leaps in computing capabilities coupled with the availability of large data sets and breakthroughs in  \emph{machine learning} (\ML)---a prominent subset of AI approaches,  which employs algorithms that  rely on a variety of methods to build machine-interpretable knowledge representations---have now ushered in a new era of innovation: One, in which \ML-based algorithmic data-driven decision-making systems (A3DMs) are assuming an increasing role in making decisions across diverse domains reaching from lending~\cite{Aziz2019}, 
insurance, and housing to college admissions~\cite{Jamison,Thomas16}, hiring~\cite{Chalfin2016}, and the justice system~\cite{Kugler2018,Heidari2019}.

On the bright side, such AI systems can easily map humanly untraceable correlations,  outperforming their human counterparts.  Better results in cancer diagnosis in healthcare~\cite{LITJENS201760} and correcting potentially substandard performance of human judges who
\begin{quote}
\emph{``effectively under-weight key observable variables like prior criminal record''}~\cite{Kleinberg2017}
\end{quote}
 are just two examples of the benefits \ML-assisted decisions can provide. 

The flip side of the coin is, however, that \ML-based AI systems are prone to learn biases present in training data and incorporate systematic error, which are unacceptable where their decisions crucially affect human lives~\cite{Osoba2017AnII}. In part, this is an inevitable by-product of their inductive-learning nature,  but there are many other sources of problems.  For instance, for various reasons, exploiting big-data sets with \ML\/ tools can lead to discriminatory models that issue unfair decisions~\cite{Corbett2018}. 
Moreover, \emph{Adversarial examples}---i.e, inputs where the \ML\/ systematically produces the wrong answer although humans would easily  perform correctly~\cite{athalye2017synthesizing,brown2017adversarial}---are an issue with deep neural network-based classifiers. Some architectures also tend to get very opaque, and it is particularly challenging to design and train systems in a manner that  aligns with human values~\cite{Gunning_Aha_2019}. 
Academic literature and a wide range of domestic and international policy documents are striving to put forward proposals to alleviate these problems:  Topics in the focus of attention include algorithmic fairness and transparency~\cite{Lee2019},  explainability~\cite{Holzinger2018,miller2019,rudin2019},
human-in-the-loop~\cite{Amershi2014,vandenelzen2011},
 human-centric and ethical design of AI systems~\cite{OECD2019,IEEE2019}, legal liability~\cite{EE2020}, and many more aspects contributing to the trustworthiness of AI technologies~\cite{Ustun2019,LD2020,EUReport,EC2020,EMCCD2020,FFLT2020,MR2020}.  

The current perspective is that AI applications---like other machines---are but mere tools lacking moral agency~\cite{SK2018,FES2020} and legal personality~\cite{CK1994}. Thus, any harm they may bring about is by definition the result of human stakeholders' poor design or misuse, for which these human actors rather than the artificial agents are responsible~\cite{SW2020}.

Still, it is interesting to look into when exactly we would deem an AI system fair or ethical. Even before the advent of AI, views on the right or ethical behavior in any given situation have been varying widely across different ethical theories, cultures, and time dimensions. Most of these differences have proven themselves to be virtually insurmountable and persist to date. Somewhat reassuringly, however, we understand the mathematical principles of AI technologies and can---as long as the prior probabilities of new cases are known and remain unchanged---accurately estimate their precision in unseen cases within the same context. But there are substantial technical limits to instilling moral values into AI systems. The challenges have been epitomized by the so-called \emph{value alignment problem} that requires AI to be deployed with moral values aligned with those of humanity. 

Various proposals have been put forward advocating the design of AI systems that are capable to reason about the virtue of  their actions~\cite{Hall} or otherwise exhibit moral agency~\cite{Adam2008,FuenmayorB20,Wallach}.  Some commentators propose to supply AI with moral-decision making ability using deontic reasoning~\cite{Young1995,BringsjordGMS18} or conceptual spaces. An example for the latter approach is the idea to construct a geometric space of moral principles and solved ethical cases and make decisions by finding the Voronoi cell for a given situation~\cite{Peterson2019}. This can be either categorized as case-based reasoning~\cite{Sriram1997} or instance-based learning~\cite{Witten}. Thus, challenges arise both when determining (1) which and how many dimensions suitably conform the moral space and (2) what specific metrics should be used to find the nearest solved moral situation or principle. These examples illustrate that, at present, technical solutions to the value alignment problem are significantly incomplete.

Others advocate that \ML\/ should incorporate equal opportunity by design. This suggestion is based on a series of proposals on various notions of fairness and/or non-discrimination such as \emph{independency constraints}~\cite{Calders2009}, \emph{group and individual fairness}~\cite{Zemel2013}, \emph{equality of  odds}~\cite{Hardt2016}, and \emph{predictive value parity}~\cite{Kleinberg2017measure}. Apart from practical obstacles of implementing such fairness criteria in the design of algorithms, this begs the question of which notion of fairness, if any, deserves priority. Again, opinions differ, with some rejecting the idea altogether, arguing that ethics always implies human choice~\cite{Vallor2018}, hence machines should never be allowed to meddle with human affairs~\cite{Rosenberg2004}. The use of AI in legal processes is also heavily criticized~\cite{Piccolo2018}. This diversity of views introduces additional layers of complexity in the debate, making the prospect of international consensus rather remote.
 
In this paper, we do not aim to include the full spectrum of AI approaches and all the possible ways in which they may affect human lives.  Instead, we only deal with \ML-based AI systems that perform supervised classification tasks.
 This is because, even before the emergence of deep-learning, most \ML\/ and data mining deployments center around classification tasks~\cite{Wutop10AI}: Fields, such as computer vision~\cite{Himanshu} and Internet of Things~\cite{Cui2018},  are dominated by applications that build and deploy classifiers. Supervised model training is the core method to master classification tasks in specific industrial settings, such as construction~\cite{HONG2020109831}. We only consider moral choices made by such AI systems, i.e., voluntary, reasoned choices that affect others~\cite{Quinn}. That said, we do not attempt to go to the bottom of centuries-long, intricate, moral debates and how these may or may not apply to \ML-based AI systems.  The reason for this is that an ethical decision---on whatever ethical basis---is only conceivable under a full-information assumption at  the time of decision-making.  However, virtually all settings in which reliance on AI decision-making is necessary are characterized by some degree of uncertainty  (the outcome is typically in the future)---indeed, the presence of uncertainty is the very reason for using AI predictive systems.  Against this background---without addressing specific moral justifications---we model moral decisions of artificial agents in a
 game-theoretic manner with the objective to arrive at two conclusions. 

Our first conclusion proposes that there exist random classifiers that perform at least as well as their deterministic piers, making them the optimal choice of classifier in some settings.  This finding is significant, as humanity's notion of \emph{automation} has so far been predominantly deterministic:  Automation has been practically equivalent to determinism, in that machines are considered to be deterministic devices implementing functions,  whereby from a given state and input only one successor state is possible, unless the machine is malfunctioning (e.g., if we push button ``5'' in an elevator, we expect it to drop us off at the 5th floor of the building,  not some unpredictable level). Similarly, the vast majority of supervised ML literature~\cite{Bishop,Hastie,Mitchell1997,Michie95} considers the search for a classifier a search for  a \emph{mathematical function}, and therefore implemented by a deterministic algorithm that consistently produces the same output from a given input~\cite{Hastie}.
\begin{quote}
\emph{``Our goal is to find a useful approximation $f'(x)$ to the function $f(x)$ that underlies the predictive relationship between the inputs and outputs''}~\cite{Hastie}.
\end{quote}
\emph{Feedback-loop controllers} are also crucially defined as functions---to each specific input, there is only one output---\cite{AstromMurray},  and robot safety has been equivalent to robots consistently keeping their distance and stopping fast enough when people are around~\cite{Malm2019}. 

For our second conclusion, we argue that in a game-theoretic model of the interactions of the decision-making AI and other rational agents,  the AI may only be able to behave optimally---i.e., so as to yield a Nash Equilibrium---if it follows a mixed strategy, that is, a randomized algorithm. This follows from the fact that games are guaranteed to have a Nash Equilibrium only in the space of mixed-strategies. Also, by equating settings with incomplete information to the game of matching pennies, we show that  any pure strategy the AI may adopt would be sub optimal. Moreover, pure strategies, when used repeatedly, can---and eventually will---be exploited by other (human) agents, which seek to selfishly maximize their own utility. Thus, mixed strategies are the most suitable strategies when decision makers face repetition (i.e.,  repeated games)~\cite{GeckilAnderson}. 

Therefore, paradoxically, it would seem that in certain situations, a truly ethical, fair, and trustworthy AI must behave randomly  and gamble with our faith, regardless of which ethical theory and/or notion of fairness it is basing its actions on. This is not to say that we are opposed to the idea that automated decision-making systems have strong ethical foundations. On the contrary, these play a crucial role in our models. However, we argue that within those moral frameworks,  there exist situations, in which fairness may dictate randomized choices.  Accordingly, the focus of this paper is not so much on any particular notion of fairness and/or ethical theory as on the conditions  under which random decisions are unavoidably necessary and what they imply. 

Accepting the premise that trustworthy AI must act randomly under specific conditions has certain profound implications. First, we will need to revise our deterministic notion of automation. Doing so will require a fundamental mindset change---be it in the general public, businesses, academia, or among policymakers. This will not be easy to achieve. That said, reassuringly, there is already some recognition that randomization cannot be simply equated with erratic, irrational, unpredictable, uncontrollable, and hence irresponsible actions, but is a very much rational---sometimes even optimal---behavior~\cite{GeckilAnderson}: 
\begin{quote}
\emph{``Game theory models, by definition, focus on the effects of one party's decision on another's interest. Such decisions are not random ones. We assert this is true even when an opponent follows a `mixed strategy' that includes the use of random actions, because the choice to use randomization as the basis for an action is a specific decision itself, not a  random event.''
}
\end{quote}
As we pointed our earlier, the field of machine ethics~\cite{Moor2006,Wallach} has also adopted that an  explicit ethical agent must display rationality by logical argumentation and justify moral decisions with some form of cognition~\cite{Serafimova}. However, others are critical of this position~\cite{Allen}. 
\begin{quote}
\emph{``What we don't want to do is make the stronger claim that the whole of cognition is just theorem proving,
 model finding, expectation maximization, or some similarly general inference procedure''}~\cite{Bello2013}.
\end{quote}
We emphasize that game theory solutions explicitly consider the rationality of all decision-makers: One player's decision process takes account of those of other players'. Rationality is thus not only relevant to game theory, but also constitutes a form of theory of mind~\cite{cuzzolin_morelli_cirstea_sahakian_2020,premack_woodruff_1978}. Such modeling of others for  moral decision-making is also present in debates on machine ethics~\cite{Clark2010,Bello2013}. To the best of our knowledge, only a few papers have discussed game theory as a mechanism for moral agency~\cite{Hall}. We explore this matter further in Subsection~\ref{TheCaseForMixedStrategies} where we examine mixed strategies. 
Colman  uses game theory to discuss Kant's categorical imperative and other ethical matters~\cite{Colman99}. This discussion shows that game theory concepts are useful to refine philosophical arguments. However, Colman's presentation fails to consider mixed strategies to debate issues such as the universality of and balance between altruist and selfish behavior~\cite{Colman99}.

Second, accepting that trustworthy AI may have to act randomly also offers an opportunity to reflect on the appropriate levels of transparency requirements for AI systems. Although random classifiers are somewhat less susceptible to exploitation, we later show that transparency may raise issues even here.

Finally, we face an implementation problem:  Under the current state of art of technology it is debatable whether true randomness can be achieved.

We report a empirical study consisting of a series of surveys to test the public's attitude towards randomized artificial decision-makers. Our results show that despite the above mentioned difficulties of implementing randomized AI systems, humans do not seem to be particularly concerned about machines randomly deciding over various aspects of their lives. On the contrary, they often prefer machine-made decisions over human decisions.

The rest of the paper is structured as follows:  Section~\ref{Prelims} presents the necessary  notations to make a case for our argument that in certain situations, AI systems need to use randomized algorithms to behave optimally. 
Section~\ref{Results} outlines our results. Section~\ref{ImplandPol} considers certain implementation and policy issues that arise from the use of random classifiers. Section~\ref{casestudy} presents our empirical study, and Section~\ref{Conclusion} concludes.

\section{Preliminaries}
\label{Prelims}

In this section, we outline some basic \ML\/ and game theory notions necessary for our purposes,  as well as notations we will use throughout this paper.  For a detailed overview of \ML\/ see~\cite{Bishop,Hastie,Mitchell1997,Witten} 
and on cooperative and non-cooperative game theory, see~\cite{Hespanha,Myerson97,rot:b:economics-computation}. 

\subsection{Machine Learning}

The most dominant application of AI technologies by several indicators and by several orders of magnitude is \ML~\cite{Columbus}. For instance, \$28.5 billion USD was invested in machine learning applications in the first calendar quarter of 2019, leading all other AI investment categories~\cite{Columbus}. \emph{Supervised learning} is the family of \ML\/ algorithms that enables  a computer system to receive labeled data and produce another algorithm---called a \emph{classifier} in case of classification tasks or more generally a \emph{model} for other tasks---that will obtain accurate predictions when supplied unlabeled  data. More precisely, assume we have a large data set $C=\{ c_{1},\ldots,c_{n}\}$ of cases. For each case $c_{i}$ we have a vector (also known as \emph{attribute or feature vector}) $\vec{x}_{i} \in \mathbb{R}^d$  of values for some properties. We also have a \emph{label} or \emph{class} $\mbox{\cal A}(c_{i})$, where $\mbox{\cal A}(c_{i})\in \{A_{1},\ldots,A_{k} \}$. If the set of possible labels is finite, we face a classification task, if it is infinite, a regression task.

The most common formulation of the \emph{induction principle}~\cite{CherkasskyM98} for supervised learning is to minimize the cost of misclassification---i.e., maximize accuracy---in \emph{unseen cases}. The latter are sometimes also referred to as \emph{cases in the future}~\cite{Hastie} or \emph{cases outside the training set} $C$. In any case, the classifier's job is to complete the missing information (provide the label)
for an unlabeled case.

Statistical decision theory~\cite{Hastie} analyzes the accuracy in the future
as follows~\cite{Michie95}\footnote{We chose an early reference on
\ML\/ terminology to ensure is compatible with all recent developments, but others~\cite{Hastie}  offer similar notation and treatment.}: We assume that there is a prior probability $\pi_j$ for each class $A_j$. That is, $\pi_j=\mbox{Prob}(A_j)$ means that the chances of running into a case whose label is $A_j$ in the future is given by $\pi_j$. The \ML\/ algorithm will build a classifier $M$ that will map the vector $\vec{x}_{c}$ (of attribute values of case $c$) into a label $M( \vec{x}_{c} ) \in  \{A_{1},\ldots,A_{k} \}$. Every time $M( \vec{x}_c ) \neq  \mbox{\cal A}(c)$, the classifier $M$ commits a misclassification error that has a cost $\kappa(c,M( \vec{x}_c ))$. If $M$ ignores all information about a case, i.e., ignores the attributes $\vec{x}_c$,
and deterministically chooses the class $A_d$ for all inputs, it incurs a cost of
\[
\sum_{j} \pi_j \kappa(c_j,d),
\]
where $c_j$ denotes a case in class $A_j$. Choosing within the family of models that ignore all evidence and information about a case can be optimized and the optimal model (under uniform miss-classification costs)  in the family responds with the class with the highest prior probability. \ML\/ generally also involves a heuristic search algorithm within a typically very large family of models to find the member of  the family that optimizes accuracy estimators and potentially other structural properties.  We usually do not know the $\pi_j$'s---otherwise there is little learning to do. 

Continuing with our discussion, \ML\/ algorithms do deliver classifiers that  examine and use the evidence, i.e., the attributes $\vec{x}_c$ of case $c$. 
The relevant probability now is the conditional probability $\mbox{Prob}(A_{j} | \vec{x}_c )$ of the class $A_j$ being the  correct class when we observe $\vec{x}_c$ in the input case $c$. Thus, to identify the model of minimum cost, this should be conditional on the information $\vec{x}$. Now, when $M$ decides $A_d$ when the evidence is $\vec{x}$, the expected cost is 
\begin{equation}
E[[M\mbox{ decides } A_d ]] =
\sum_{j} \mbox{Prob}(A_j | x) \kappa(j,d), \label{costRaw}
\end{equation}
where $\kappa(j,d)$ is the cost of miss classifying as $A_d$ when the case actually  belongs to $A_j$. Using Bayes theorem, 
\begin{equation}
 \mbox{Prob}(A_j | x) = \pi_j \frac{ \mbox{Prob} (  x | A_j) } {\sum_s \pi_s  \mbox{Prob}(x | A_s) }. \label{Bayes}
\end{equation}
However, since $ \sum_s \pi_s \mbox{Prob}(x | A_s)$ is a constant, we can simplify Equation~(\ref{Bayes}) and then replace $\mbox{Prob}(A_j | x)$ in Equation~(\ref{costRaw}). We conclude that deciding on $A_d$ has a cost of
\begin{equation}
\sum_{j} \pi_j \kappa(j,d)  \mbox{Prob} (  x | A_j),
\end{equation}
modulo a constant factor that is the same for all $d\in \{1,\ldots,k\}$. Under the assumption that these are continuous distributions where the probability density function is $f_{j}(\vec{x})= f(\vec{x} | A_j)$, the optimum is the $d\in \{1,\ldots,k\}$ where
\begin{equation}
\sum_{j} \pi_j \kappa(j,d)  f_{j} ( \vec{x}) \label{Distributions}
\end{equation}
is minimum. In the case of discriminating between two classes $j_1$ and $j_2$ ($k=2$) and $\kappa(j_1,j_1)=\kappa(j_2,j_2)=0$, Equation~(\ref{Distributions}) reduces to $M$ declaring class $A_{j_1}$ if
\[
\frac{ f_{j_1}( \vec{x} ) }{f_{j_2}( \vec{x}) } > \frac{ \pi_{j_2} \kappa(j_1,j_2) } { \pi_{j_1} \kappa(j_2,j_1)}.
\]
The classifier $M$ should declare $A_{j_2}$ otherwise. This is a fundamental likelihood ratio all supervised \ML\/  algorithms are attempting to optimize---commonly even assuming $\kappa(j_1,j_2)= \kappa(j_2,j_1)$.

We now interpret this result examining some simple cases. Consider not only $k=2$ but $\kappa(j_1,j_2)=1$ when $j_1 \neq j_2$. Moreover, we will assume $ \pi_{j_1}= \pi_{j_2} =1/2$. That is, we have a balanced mixture from where our cases are drawn, and with equal probability we have a case from class $A_{j_1}$ or class $A_{j_2}$. If we do not look at the evidence, choosing $M$ as a constant function will make errors $50\%$ of the time (the two classifiers in the family of constant functions are as good or as bad).

If the classifier can examine the evidence $\vec{x}$ and we know $f_{j_1}( \vec{x})$ and  $f_{j_2}( \vec{x})$, the optimal classifier is given by
\begin{equation}
M( \vec{x} ) =
\left\{ 
\begin{array}{ll}
A_{j_1} & \mbox{if } f_{j_1}( \vec{x}) > f_{j_2}( \vec{x}), \\
A_{j_2} & \mbox{if } f_{j_2}( \vec{x}) > f_{j_1}( \vec{x}). 
\end{array}
\right.
\end{equation}
This should be intuitively clear. In fact this leads to \emph{Fisher's Linear Discriminant} being the optimal classifier if the environment draws classes from a mixture of two multivariate normal distributions with
\begin{enumerate}
 \item equal diagonal covariance matrices $\Sigma = \lambda I_{d}$ and 
\item
 equal weights, i.e., $1/2$ each. 
\end{enumerate}
If we have a mixture of more than two classes  but all the weights equal to $1/k$ and all covariance matrices are again diagonal matrices with the value $\lambda$ in the diagonal, the optimal classifier is the $k$-nearest neighbor
that, on vector $\vec{x}$, declares the class where the mean $\vec{\mu}_{j}$ is nearest to  $\vec{x}$. One way to visualize this last case is to note that the multidimensional bell shape of the multivariate normal 
\[
\mbox{\cal N}(\mu,\Sigma) = (2 \pi)^{-d/2} det( \Sigma)^{-1/2} e^{-\frac{1}{2} (\vec{x}-\vec{\mu})^{T} \Sigma^{-1} (\vec{x}-\vec{\mu})}
\]
will have level curves that are spheres around $\mu$ and $f_{j_s}( \vec{x}) > f_{j_t} ( \vec{x})$ if  $\vec{x}$ is closer to $\vec{\mu}_{j_s}$ than to $\vec{\mu}_{j_t}$. Note that when $\Sigma = \lambda I_{d}$ for each class, $(\vec{x}-\vec{\mu})^{T} \Sigma^{-1} (\vec{x}-\vec{\mu})$ is the Euclidean metric scaled by the same factor for all components of the mixture.

\subsection{Game theory}

At present, the vast majority of artificial decision-makers operating in all sorts of industrial settings are constructed by applying supervised \ML. There are huge data sets of labeled cases and digital records of attributes. The classes are decisions---for instance whether or not to award credit to a customer based on the data indicated in their credit application. Some of these decisions have ethical consequences because they impact individuals' well-being.
At the extreme, classifiers as autonomous machines may have to make life-and-death decisions (e.g., targeting decisions of autonomous weapon systems and \emph{trolley problem} type of scenarios~\cite{Foot1967}, where the classifier has to decide whether to harm individual(s) $X$'s or individual(s) $Y$'s welfare). In such cases, it is of paramount importance to make the right decision, so the question arises: How can such algorithms make optimal choices? Game theory---which studies conflict and cooperation between rational agents---is helpful to answer that question.  

A \emph{game} abstracts the fundamentals involved in decision-making---such as agents' possible moves, actions, choices, and decisions---and formalizes \emph{rationality} as acting to maximize \emph{utility}. Games can be either cooperative---where agents form factions in order to maximize their utility---or non-cooperative---where agents do not coordinate their actions.  This paper only uses non-cooperative settings where one agent is the \emph{environment} (or \emph{future}), which clearly cannot cooperate. It is common to apply game theory with such an adversary to achieve \emph{robust designs}~\cite{Hespanha}.

Given a game, a solution is to determine the most rational decision each agent should reach.  Importantly, all information for making the decision is represented in the modeling of  the game---if there was more information, we would be discussing a different game. The game also identifies the decision-makers or influencers as \emph{players} and all options available to them---typically a set of exclusive decisions.  Formally, in a game with $p$ agents, for each agent $P_i$, $1\leq i \leq p$, there is a set $D_i$   that defines agent $i$'s set of possible decisions it can pick.  The elements of this universe of decisions are called \emph{pure strategies}. The solution of the game is the set of strategies that each agent must perform to maximize their individual utility. In a game with $p$ players, the value each player will receive when all players have declared their decision is specified  ahead by \emph{payoff functions} (sometimes also called the \emph{payoff matrix}).  These are typically a set of functions $u_{j}(d_{1},\ldots,d_{p})$ that  provide the utility to player $P_j$, $j \in \{1,\ldots,p\}$, when each player $P_i$ makes decision $d_{i}$, for all $i\in\{1,\ldots,p\}$.

From very early on in the study of game theory, it was observed that players can also use \emph{mixed strategies}.  A mixed strategy is a randomized algorithm, i.e., a probability distribution over the possible pure strategies.  A \emph{Nash equilibrium} is a solution that suggests the ultimate rational behavior for agents: It implies that---even if other players' decision-making strategies became known to them---none  of the players $P_i$ would have any incentive to change their behavior.  That is, in a Nash equilibrium, every player selects its strategy and cannot do better even if other players disclose theirs: For every player $P_j$, despite knowing the choice $d_{i}$ for all other players $j\neq i$, changing their choice $d_j$ cannot increase the value of  $u_{j}(d_{1},\ldots,d_{p})$. An alternative characterization is that, in a Nash equilibrium,  a player's choice is the best response to the others' best-responses (and this is true for each player). While every finite game has a Nash equilibrium with mixed-strategies, a Nash equilibrium with pure-strategies may not exist. 

A very famous illustration of game theory applicable to many cultures is the game \emph{rock, paper, scissors}. Here, each player makes a decision selecting one from the set $D= \{ \mbox{\emph{rock}},  \mbox{\emph{paper}}, \mbox{\emph{scissors}} \}$. The choices are simultaneous. \emph{Scissors} cut \emph{paper}, \emph{rock} breaks \emph{scissors}, and \emph{paper} covers \emph{rock}. Using the pure mathematical formulation of the game, it is clear that programming a machine to play this game deterministically is vulnerable to exploitation after inspection---there is no Nash equilibrium with pure strategies. However, a mixed strategy---a randomized algorithm that selects one of the three pure strategies with probability $1/3$---will win the same number of times as it ties or loses. This results in zero expected utility, which happens to be the optimum.  It follows that, facing an opponent using a randomized strategy, it is best to use the same mixed-strategy as the opponent.

\subsection{Equal Opportunity Principle}
  
As emphasized earlier, for artificial decision-makers to be trustworthy and fair,  they must base their decisions on some ethical theory and/or notion of fairness.  When learning a classifier to enable their decision-making capability, AI systems need to convert some ethical principle(s) and/or fairness notion(s) into a formulation of the costs $\kappa$ of miss-classification.  Similarly, the payoff functions of games modeling the interactions of the decision-making AI and its environment are ideally defined based on  some ethical principle(s) and/or fairness notion(s).  In line with our focus on the necessity and implications of randomized classifiers that employ mixed strategies,  rather than debating the vices and virtues of different fairness concepts and ethical theories, we will adopt the principle of \emph{equality of opportunity (EOP)}.

The reason for our choice is that the EOP is a central ideal of fairness studied in political philosophy and economics,  which has been suggested to be a generalization of several measures of fairness for \ML~\cite{Heidari2019}. We remark again that \emph{statistical parity}~\cite{Feldman2015}, \emph{equality of  odds}~\cite{Hardt2016} \emph{equality of accuracy}~\cite{BuolamwiniG18}, and  \emph{predictive value parity}~\cite{Kleinberg2017measure}  can be regarded as particular forms of EOP~\cite{Heidari2019}.  Essentially, the EOP aims to establish an initial level playing field---equality of opportunity---among individuals based on their innate qualifications  rather than exogenous factors.   At the same time, it allows for competition between individuals for different positions. This may result in  different \emph{outcomes}, which are morally acceptable as long as  they reflect individuals' \emph{merit} or \emph{deservingness}.  Thus, the EOP distinguishes between morally acceptable and unacceptable inequality. 

Note that merit is highly contested~\cite{McCRUDDEN} and many question whether providing true equal opportunity is feasible in practice~\cite{Fishkin}. These concerns are particularly relevant for AI systems, which typically perform moral decisions based on incomplete information:  With all likelihood, they do not consider all attributes of merit or measure merit accurately,  all of which impairs their ability to precisely ascertain the differences in opportunity. 
 
But we need not deal with these problems here: What matters for our argument that AI decisions must use randomization is not \emph{what} particular ethical principle/fairness notion we use, but rather \emph{that} we need to use one of them. This is because, reliance on some ethical principle---in our case the EOP---is instrumental to, given a case $c$, defining $d$ attributes and their values that define the vector $\vec{x}_{c}\in \mathbb{R}^{d}$ for the AI classifier to make decision $M( \vec{x}_{c})$. Consider the example of a sentencing decision made by an AI system. In deciding the length of time in prison, merit attributes would be considered: What degree of remorse has the offender shown,  how did their actions impact the victim(s), was the offense intentional and premeditated or rather reactive and emotional?  These would constitute attributes of merit, ultimately determining the length of the sentence. Alternatively, in an AI-assisted parole hearing or release assessment, the application of the EOP may lead to considering attributes of (positive) merit like good behavior and increased educational level. These examples illustrate the role of ethical principles, such as the EOP, in dictating the dimensions (attributes) of the vector for an AI-based system to make a decision.
They also give a feel for what we mean by incomplete information: At the time of sentencing, the AI-judge has to make a decision in the absence of the valuable hindsight that is available from the data to the AI-judge at the time of the parole hearing.

In Appendix~\ref{EOPintro}, we reproduce the instructions to the subjects of our empirical study and the background on the EOP that was supplied to them.

\section{The Case for Non-Deterministic Classifiers}
\label{Results}

Although randomization (non-determinism) is prevalent in mathematics,  computer science, and artificial intelligence, we believe that the focus on deterministic classifiers derives from the tradition to perform the analysis of misclassification error using statistical decision theory~\cite{Hastie}.
As we already explained, such analysis formulates that classifiers should be mathematical functions, selected by the learning process from families of functions~\cite{CherkasskyM98}, and thus deterministic mappings from inputs to outputs reflecting the \emph{decision boundaries}~\cite{HandSmythMannila} in the classes.  This view of classifiers as functions reflecting decision boundaries~\cite{DudaHart} or decision surfaces~\cite{Bishop} is pervasive across models, whether it is artificial neural networks~\cite{Beale1990NeuralC}, support vector machines~\cite[Page~2]{cristianini_shawe-taylor_2000}, conceptual learning~\cite{Langley}, or decision-tree learning~\cite{Thornton}. 

\subsection{Random Classifiers May Be Optimal}
\label{RandomCassifiers}

Despite the predominantly deterministic approach towards automation so far, we argue that randomized algorithms~\cite{MotwaniR97} are mathematically valid families from which to infer and implement classifiers.  We describe a simple case in which a randomized algorithm is optimal, i.e., no \ML\/ algorithm can produce a deterministic classifier with less misclassification cost.

\begin{proposition}
There exist settings where randomized classifiers are optimal, i.e., no other classifier learned by any supervised learning technique can deliver a classifier with less misclassification cost. 
\end{proposition}

\begin{proofs}
Consider a scenario where the evidence is just one value $x \in \mathbb{R}$ (i.e., a vector of dimension $d=1$) and where the distribution of the two classes $f_{j_1}( x)$ and $f_{j_2}( x)$ is given as follows. Let the probability distribution of the first class be uniform on the interval $[0,b]$: 
\[
f_{j_1}( x) =
\left\{
\begin{array}{ll}
\frac{1}{b} & \mbox{if } 0\leq b \\
0 & \mbox{otherwise}.
\end{array}
\right.
\]
The probability distribution of the second class is also uniform but on the interval $[a,c]$ of the same length:
\[
f_{j_2}( x) =
\left\{
\begin{array}{ll}
\frac{1}{b} & \mbox{if } a\leq c \\
0 & \mbox{otherwise},
\end{array}
\right.
\]
with $0<a,b<c$ and $b=c-a.$ Moreover, we assume the mixture of these two distributions has equal weights with value $1/2$. See Figure~\ref{MixtureOfUniformDistributions}.
\begin{figure}
\centering
\includegraphics[width=0.5\textwidth]{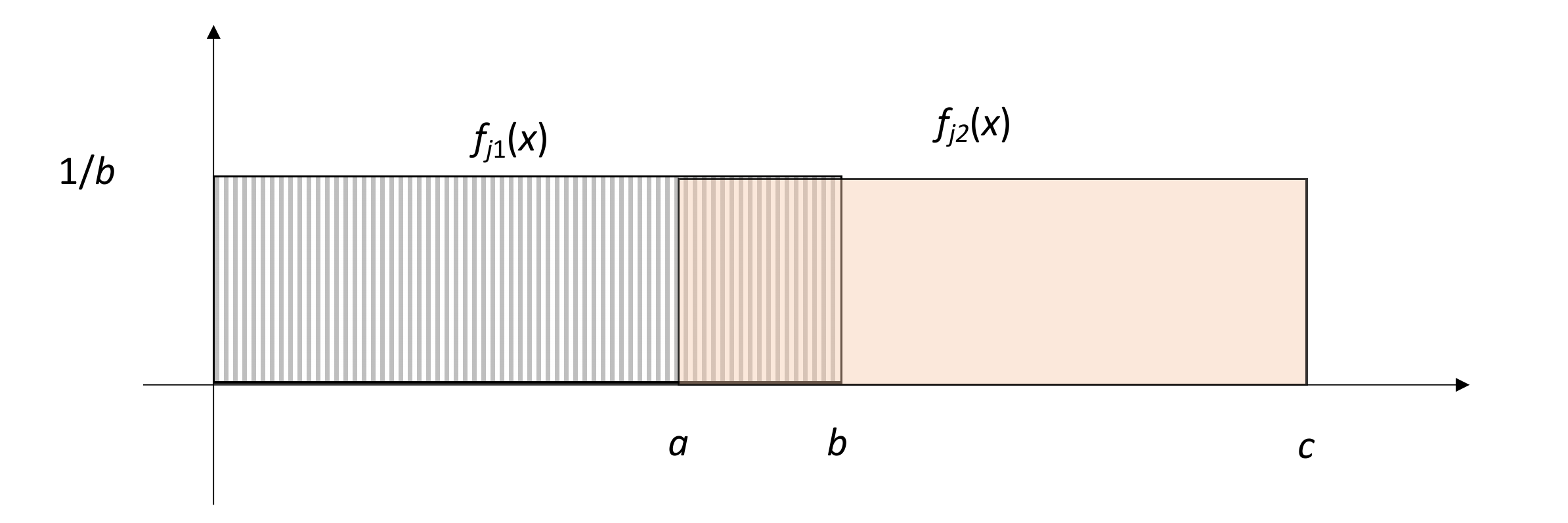}
\caption{\label{MixtureOfUniformDistributions} Two uniform distributions with support of length $b$, where the second is the first shifted by $a$.}
\end{figure}

From our earlier discussion, the following deterministic classifier is optimal:
\begin{equation}
M_d (x) =
\left\{
\begin{array}{ll}
A_{j_2} & \mbox{if } (a+b)/2 \leq x \\
A_{j_1} & \mbox{if } (a+b)/2 > x. \\
\end{array}
\right.
\end{equation}
This classifier is optimal whether $a<b$, $a=b$ or $b<a$. It always chooses the higher (or not lower) between $f_{j_1}( x)$ and $ f_{j_2}( x)$. However, consider the case illustrated in  Figure~\ref{MixtureOfUniformDistributions} or the case $a=b$, and the following randomized classifier:
\begin{equation}
M_r (x) =
\left\{
\begin{array}{ll}
A_{j_2} & \mbox{if } b < x \\
A_{j_1} & \mbox{if }  x < a \\
B(1,1/2)[ A_{j_1}, A_{j_2}] & \mbox{if } a \leq  x \leq b, \\
\end{array}
\right.
\end{equation}
where $B(1,1/2)[u,v]$ is one Bernoulli trial with probability $1/2$ where $u$ is
selected on success and $v$ on fail.

We argue that $M_r$ is optimal for all cases,  even $a<b$ because by vacuity, the third case part of the definition of $M_r$ would not apply, so $M_r$ coincides with $M_d$.

If  Figure~\ref{MixtureOfUniformDistributions} or $a=b$ applies, any case with evidence $x$ with $a \leq  x \leq b$ has equal probability of coming from class $A_{j_1}$ or class $A_{j_2}$. When $a \leq  x \leq b$, the evidence is completely uninformative. If we choose the deterministic classifier $M_d (x) $ that always responds with $A_{j_1}$, it will be correct and wrong, respectively, $50\%$ of the time. The same holds if we choose $A_{j_2}$ for all $a \leq  x \leq b$. Classifier $M_d (x) $ is not constant. It makes an error  $50\%$ of the time for $a \leq  x (a+b)/2$ because it always declares $A_{j_1}$, even though $50\%$ of the cases are from $A_{j_2}$. But it also has a $50\%$ error rate for $(a+b)/2 < x < b$, declaring $A_{j_2}$ on cases that are $A_{j_1}$.

We now show that the random classifier $M_r (x)$ also makes the same expected error. Note that this analysis is the same as the profit analysis for the 2-player game \emph{Matching Pennies} introduced in the following subsection. The environment can draw from class $A_{j_1}$ or class $A_{j_2}$ with equal probability $p=1/2$. $M_r (x)$ declares  $A_{j_1}$ or class $A_{j_2}$ with equal probability $p=1/2$ as well.  The scenario is equivalent to tossing two fair coins as independent Bernoulli events, both with probability $1/2$, and declaring success when the coins match (i.e., outcomes (head, head) and (tails, tails)) and declaring a failure when the coins do not match (i.e., outcome (head, tails)). Thus, classifier $M_r (x)$ is correct $50\%$ of the time and incorrect $50\%$ of the time.
\end{proofs}

We may be tempted to say that for evidence $\vec{x}$ where $f_{j_1}(\vec{x}) =  f_{j_2}(\vec{x})$ \emph{it does not matter} what the classifier determines as a class. The point we want to emphasize here is that randomized algorithms/classifiers (choosing differently each time they run) are not less optimal compared to systematic (deterministic) classifiers, hence their use in artificial decision-makers is worth consideration. 

\subsection{The Problem of Deterministic Classifiers}

Matching Pennies is a simple game in basic game theory, which starts by two players submitting a covered penny, respectively, either heads or tails up.
The players then simultaneously reveal their pennies with one of the following two outcomes: (1) if the pennies match---both heads or both tails---the first player (named the \emph{even player}) collects both as profit, (2) if they do not match---one head the other tail---the second player collects both pennies. Matching Pennies is one of the simplest examples of a game where Nash equilibria only exist with mixed strategies.

We now use matching pennies to elaborate on our previous example of an AI system determining a sentence length. Aside from punishing offenders, deterring future crimes, and serving justice for victims and society, sentencing also pursues rehabilitation objectives: The intention is to calibrate sentences so as to ensure that offenders can redeem themselves and no longer pose danger to society after their release. That is, we want the AI prediction to achieve accuracy in the future. The problem is, however, that merit attributes relevant for determining whether or not the offender can be safely released into society upon completion of their sentence only materialize in the future, but none of them are available to our sentencing AI. The sentencing AI cannot possibly know if all convicts have equal opportunity to redeem themselves regardless of which correctional facility they have served their sentence in (probably not). 
This means that the vector $\vec{x}_{c}$ on which the sentencing AI supports its decision is only a projection to the present and is missing all the attributes of the future that will be known when deciding on the release. Accordingly, penitentiary systems mostly treat original sentences as mere estimates and regularly review release conditions before and during completion of the sentence. The environment plays as an adversary---as mentioned before, we alluded to this when we mentioned that game theory is used for robust design~\cite{Hespanha}. Thus, the sentencing AI is playing the game of matching pennies with the penitentiary system as to whether the sentence will
achieve all its objectives.

Much like this example, moral decisions made by AI systems are characterized by incomplete information. Often, cases look equivalent in the current set of attributes (in the current projection), but they are missing information on attribute values that are only completed later on. Suppose we are to construct a classifier that makes decisions affecting the well-being of two individuals: For instance, amidst the COVID-19 pandemic, a hospital has only one intensive care unit (ICU) available and two patients ($c_1$ and $c_2$) require it. We wish to allocate the ICU to the patient with the highest probability of survival.
We feed extensive clinical and virological data~\cite{Chen2020,Lescure2020,ZhouLANCET2020}  of identified risk factors from the now millions of patients infected with the disease into ML algorithms to obtain the classifier that decides between $c_1$ and $c_2$. As per the previous section, this means we have attribute value vectors $\vec{x_1}$ and $\vec{x_2}$ that codify the values in identified risk factors (e.g., age, obesity, diabetes, to name just a few). Note that problems arise when the available feature vector is uninformative, i.e., it does not allow to distinguish between the merit of $c_1$ and that of $c_2$---they are of the same age, weight, and so on. To whom should the classifier award the ICU in such cases? Deterministically, always on a first-come-first-serve basis? Is this fair or ethical by some metric? How will the public react if it notices that the individual who is entered as first patient into the classifier's input file always receives the ICU when both patients have the same merit? 

Now consider a different example: Training an ML system to make decisions in autonomous vehicles (AVs). Suppose that---when facing an imminent collision with either a motorcyclist wearing a helmet or a motorcyclist not wearing one---vehicles equipped with such classifier select the motorcyclist with a helmet as the victim due to the lesser likelihood of sustaining serious injuries. It is not hard to see how this may eventually motivate motorcyclists to drive without helmet to avoid being selected as victims by AVs---an outcome that is clearly undesirable from a general safety perspective. A potential solution could be for the AI to play a (finitely or infinitely) \emph{repeated game}~\cite{Dutta}. For now, we will not further entertain game theoretic repeated games. Suffice to say that automating with an AI implies the use of a large amount of instances and repeated games typically result in extremely different behaviors from single interaction games. 

The list of examples could go on. The point we would like to emphasize is that individuals are likely to change their choices to exploit the classifier if its decisions are deterministic---a behavior that becomes easier the more transparent the classifier is. This shows that (1) transparency may have substantial adverse effects in pure-strategy game settings---although, as we will explain below, it poses issues even in mixed-strategy game settings, (2) deterministic classifiers cannot necessarily achieve a Nash equilibrium, i.e., behave optimally, and (3) it is of crucial significance whether or not we can anticipate all the factors relevant to making a decision, i.e., whether or not we have a complete feature vector for a particular case.

\subsection{The case for mixed strategies}
\label{TheCaseForMixedStrategies}

Our arguments in the previous sections already support the idea that AI-classifiers should use a mixed strategy and make random choices (but note that `random choice' under a mixed strategy is not any arbitrary unjustifiable choice,  it is to wisely select a particular optimal and fixed probability  distribution, which can be justified in a game theory sense, and from there to apply it to select  a pure strategy). To stress this point further, we now use a 2-player game to examine the challenge faced by a classifier that can choose only between two options. The classifier is the first player, and has two \emph{pure} strategies at its disposal: (1)~to harm the welfare of individual(s) $X$ or (2)~to harm the welfare of individual(s) $Y$. The other player is again the environment, which can place the negative merit that leads to the scenario of harm either on~$X$ or~$Y$.

We start with a scenario, where the AI faces the rare luxury that merit is ascertainable. As shown below, even such settings can raise complex moral dilemmas: 
\begin{center}
\fbox{ \parbox{0.95\textwidth}{
{\scriptsize
Two passengers in an AV are driving safely and obeying the speed limit in a lane labeled as only for AVs. As an extra safety measure, the road also has barriers to minimize pedestrians accidentally invading the designated AV lane. Six intoxicated pedestrians who have illegally jumped the barriers and failed to check if there were oncoming vehicles suddenly jump in front of the AV. Should the AV harm the pedestrians or swerve and harm the passengers? 
}}}
\end{center}
Here, the AI can ascertain that the illegally crossing pedestrians are responsible for the now unavoidable harm to human lives. Best efforts to apply brakes, yet not changing direction and causing some unavoidable harm is widely regarded as acceptable even for human drivers~\cite{Bonnefon,Estivill-CastroEthics,Moore}. Presumably, thus, such a merit-based, deterministic decision would not raise any eyebrows even in the case of AI decision-makers. That said, the best course of action dictated by utilitarian ethics may change depending on the particular circumstances of a given scenario and the \emph{ethical} action may not be \emph{feasible} in practice: For instance, the vast majority of society would accept utilitarian cars and confirm that if the six pedestrians were correctly crossing the street, the ethical decision by the AV---by utilitarian standards---would be to swerve risking the well-being of the two passengers. But, people also admit they would not own such a vehicle~\cite{Bonnefon,Estivill-CastroEthics,Moore} and would game the system by artificially adding passengers to their cars. Such practical issues may be at least to some extent alleviated if some degree of randomization is introduced into the classifier. Leaving aside the question of how exactly such randomization is achieved for the moment, we restrict ourselves to point out that this observation provides a strong argument for randomized classifiers even in scenarios where merit is ascertainable. 

Now consider a variation of this scenario where merit is not readily ascertainable:
\begin{center}
\fbox{ \parbox{0.95\textwidth}{
{\scriptsize
The two passengers in the AV are driving safely and obeying the speed limit in the same AV lane as in the previous scenario. The six pedestrians suddenly appear in front of the AV. However, they are neither intoxicated nor crossing the road unlawfully. Rather, they cross over a zebra-indicated zone that is also equipped with traffic lights, which unfortunately do not work. It is unknown, whether or not they took all reasonable safety measures before crossing. What should the software be programmed to do?
}}}
\end{center}
In this new scenario, it is likely that the AI---while reasoning about merit---reaches the conclusion that it is not ascertainable. We already mentioned that many oppose against any machine participation in moral decisions impacting humans. In the case of AVs, this is confirmed by the Report of the German Ethics Commission on Automated and Connected Driving~\cite{GERMANreport2017}, suggesting that any tragedy involving AV's would be attributed to humans only and not machines. Thus, authorities regard our current AI systems as fully the responsibility of their human designers---a premise that implies the following reasoning by the AI:

\begin{enumerate}
\item Only human agents should decide, but I am facing a decision to choose between the welfare of one human being over that of another.
\item This situation should not have occurred.
\item Therefore, some human agent is at fault.
\item  According to the EOP, I must optimize for the number of human casualties and, in particular, the consequences of this tragedy should be endured by  the human agent(s) at fault.
\end{enumerate}
However, the same report~\cite{GERMANreport2017} admits the AV would not be in a position to fully ascertain which human agent is at fault or the level of negative merit---the gravity of fault---at the time of decision-making, necessitating settlement of the matter before court. 

The challenges of this second scenario, where the AI cannot determine merit and must choose between two evils---harm individual(s) $X$'s or individual(s) $Y$'s welfare---can be formulated in a game analogous to matching pennies and resolved with mixed strategies as follows: For simplicity, we assume---like others~\cite{Bonnefon}---there is no other information. Also, we model merit as negative utility. One player is the artificial decision-maker, which has only two \emph{pure} strategies at its disposal: (1) to harm the welfare of individual(s) $X$ or (2) to harm the welfare of individual(s) $Y$. The other player is the environment, which can place the negative merit that leads to the scenario of harm either on human $X$ or on human $Y$. In the scenario involving only two humans, the payoff matrix for the artificial agent is as follows:

\begin{center}
\begin{tabular}{l  c  c } 
 & \multicolumn{2}{c}{\tiny Payoff matrix for the AV} \\
 & \parbox{2cm}{\tiny Human~$X$ was at fault} 
 & \parbox{2cm}{\tiny Human~$Y$ was at fault}  \\  \cline{2-3}
\parbox{2cm}{\tiny AI chooses to harm~X} & \multicolumn{1}{|c|}{0} & \multicolumn{1}{c|}{$-m_{X}*v_{X}$} \\ \cline{2-3}
\parbox{2cm}{\tiny AI chooses to harm~Y} & \multicolumn{1}{|c|}{$-m_{Y}*v_{Y}$} &  \multicolumn{1}{c|}{0}  \\ \cline{2-3}
\end{tabular}
\end{center}

When the AI chooses to harm the human who is at fault in causing the scenario, there is no penalty (no negative utility). But when it chooses to harm the welfare of the innocent human, the AI pays the price of the merit of the negatively affected human times a value of worth for that human---this allows us to account for several pedestrians or several passengers in our illustration.
So, harming $X$ when the fault was $Y$'s impacts negatively on humanity in the amount of the merit of $X$ times the number of humans of type $X$. The payoff matrix for the environment which can choose to cause a scenario with $X$ at fault or $Y$ at fault is the same, except the values are positive (it is a zero-sum game).

For this game, it is not hard to show that there is no Nash equilibrium with pure strategies, and that in a Nash equilibrium with mixed strategies the artificial agent chooses to harm individual(s) $X$ with probability
$m_{Y}*v_{Y}/(m_{X}*v_{X}+m_{Y}*v_{Y})$ and individual(s) $Y$ with probability
$m_{X}*v_{X}/(m_{X}*v_{X}+m_{Y}*v_{Y})$~\cite{vonStengel02}.

The exact probabilities are not crucial for our argument here. The important message is that the algorithm is randomized. If the AV cannot be sure who is at fault, its two passengers or the 6 pedestrians, then despite the fact that six pedestrians are more lives than the two passengers, the EOP commands that the passengers still deserve a roll at the die---with probability $2/8$---to be saved! This illustrates our point that whenever an AI decision-maker does not possess full information on merit---as is the case in the vast majority of settings in which we engage AI systems---randomized classifiers constitute the optimal and fairest option.

David McFarland's \emph{Epilogue: The Alien Mind} uses ethical dilemmas to debate the characteristics and properties we would need to anticipate  on an alien mind to qualify it as one of us (among them, the robot would need to be accountable  for its actions and provide a rational justification for its actions)~\cite{McFarland}. Being amoral is not an argument against random classifiers,  as the law deals with other instances of ``amorality''' like insanity~\cite{Bonnie}. Moreover, we already argued that current machines are regarded as amoral, whether deterministic or not.

Another important point for randomized algorithms: Most \ML\/ formulations are optimization problems where parameters of a model in a family of models are being sought for some  inductive principle~\cite{CherkasskyM98}. Such optimizations can be performed by hill-climbing, gradient descent or other heuristics, because finding the optimal model (i.e., the global optimum) is typically computationally infeasible. For instance, the famous back-propagation algorithm of artificial neural networks is a gradient descent method, and the initial weights are typically randomly chosen values.  Thus, algorithms, such as hill-climbing or gradient descent, must be randomized regarding their starting point, and robustness is achieved by taking the best model out of the randomized multi-start. Another aspect that makes supervised learning randomized is the shuffling of the dataset instances to avoid bias to presentation order.  This observation, that the learning involves randomization,  applies to supervised learning, but also to  unsupervised learning---recall the most popular of all: $k$-means starts with a random clustering. Today randomized algorithms are the dominant approach used to  build AI-models. Thus, at each successive learning occurrence, a new convolutional neural network---or any other classifier---despite being deterministic, would still be likely to produce different outputs from the previous learning occurrence. Overall, it seems somewhat contradictory to demand that the output of learning be a deterministic algorithm while what is used to obtain such  classifier is a randomized algorithm.

Our approach does not oppose the accurate representation, computer-mechanization,  and artificial reasoning with ethical theories. The adoption of moral facets amenable to logic programming~\cite{FuenmayorB20,PereiraSaptawijaya} emerges from the focus on moral  decisions being the result of reasoning. Among those adopted are the \emph{Doctrine of Double Effects} (DDE)~\cite{sep-double-effect} and \emph{contractualism}~\cite{Scanlon}. Computational versions of the Doctrine of Double Effects have been implemented with systems such as QUAL, ACONDA, and Probabilistic EPA (as well as other more recent proposals~\cite{GovindarajuluB17}). These systems have been illustrated with several versions of the trolley dilemma~\cite{PereiraSaptawijaya}. They are plagued with limitations~\cite{GovindarajuluB17}---indeed, the original source for trolley dilemmas~\cite{Foot1967} is also a critique to the DDE. Moreover, Foot presents examples where refinements of the DDE (as to whether
direct or oblique intention, or the distinction between avoiding injury and bringing aid) are insufficient to make a choice~\cite{Foot1967}. For instance,  
\begin{quote}
\emph{``shipwrecked mariners who believed that they must throw someone overboard if their boat was not to founder in a storm, [sic] there is no conflict of interests so far as the decision to act is concerned; only in deciding whom to save.''}~\cite{Foot1967}.
\end{quote}
But if all sailors have equal merit to survive; a random decision might be justifiable.

Similarly, there has been remarkable progress  towards explicit ethical artificial agents  by showing that Gewirth's ``Principle of Generic Consistency (PGC)'' \cite{Gewirth} can be derived by computer-mechanization using higher order logic~\cite{FuenmayorPRICAI2019}.
That is, from premises such as ``I am a prospective purposive agent (PPA)'' (that is, I act voluntarily for some (freely chosen) purpose $E$) and ``My freedom and well-being (FWB) are generically necessary conditions of my agency,'' the artificial agent concludes that ``Every PPA has a right to claim their FWB.'' As an ethical principle, the PGC seems to be a variation of the
\emph{golden rule}: ``treat others as you wish to be treated.'' The PGC is not uncontroversial~\cite{Beyleveld1991}.  But it has been praised~\cite{Kornai} for being instrumental to impose morality on artificial agents who would respect humans: Artificial agents with sound reasoning mechanisms would conclude  they must respect the well being of all PPAs and would recognize both humans and themselves as PPAs. But, Trolley problems are not about whether an agent is a PPA or not, but about choosing the less evil among PPAs. Moreover, such an artificial PPA endowed with morality based on deductive-reasoning ---when faced with the choice between acting against the well-being of one PPA or another---must conclude that one of them is not a PPA (particularly if we insist that AI should not make moral decisions). A further complication is that the AI performing such deductive-reasoning may still be required to make a choice
without being able to ascertain which one is or is not a PPA.

A final point in favour of mixed strategies is that they offer an interpretation of probabilities as frequencies (proportions)~\cite{Hespanha} as they apply to populations. Suppose we need to program robots for a tournament involving our earlier example of \emph{rock, paper, scissors}, where robots are only allowed to use pure strategies.  The teams consist of a large number of robots (say $1,000$), each of which we can program differently and shuffle across teams.   
The mixed-strategy solution in this situation is to program one third of the robots in each team to deterministically choose one of the three actions.
However, even this use of deterministic programs shuffled randomly is less acceptable than randomized algorithms: We would need to hide away what deterministic algorithm each AI decision-maker is using (i.e. block all transparency). Humans would seek to breach this by discovering or identifying the deterministic AI decision-maker that favors their situation above other humans.

\section{Implementation and Policy Issues}
\label{ImplandPol}

Section~\ref{Results} provided compelling arguments for employing randomized classifiers in the  above outlined scenarios. However, as theoretic arguments are not necessarily easily implemented in practice, we now look into implementation issues. 

First, the success of the societal adoption of any novel technology---here deploying trustworthy and fair  randomized classifiers to make moral decisions impacting the welfare of human beings---hinges on society's trust  and acceptance. Since humanity's notion of automation has so far been predominantly deterministic and change is often  met with considerable resistance, building societal confidence in non-determinism may not be so easy to achieve. In particular, for game theory, mixed strategies---although extremely intuitive in settings  that can be modeled by matching pennies or rock, paper, scissors---have received heavy opposition. For instance, preferential voting schemes that have been shown to  be optimal  by a game-theoretic analysis and also superior empirically have been rejected~\cite{RivestCOMSOC}. 

To assess societal preferences, we conducted a series of surveys using classic ethical dilemmas to evaluate the public's stance on randomness.  We also explored under which circumstances people prefer moral decisions to be made by humans, artificial agents, or a combination of both such that human decision-makers are assisted  by computer technology to varying degrees.  Our goal was to gather empirical evidence to help identify areas where emerging AI capabilities  could realistically automate decisions that affect human lives. 
Our experimental setting is introduced in Section~\ref{casestudy}.  We found that society would be inclined to accept random choices by artificial agents 
and feel comfortable with technology eventually fully  replacing human decision-makers in specific areas.

Second, initiatives to achieve ever-higher levels of transparency, 
traceability, and auditability of AI systems are now ubiquitous. Prominent examples include the OECD Principles on AI~\cite{OECD2019}, the European proposal for an Artificial Intelligence Act ~\cite{AIA}, the Feasibility Study of the Council of Europe's Ad-Hoc Committee on Artificial Intelligence (CAHAI)~\cite{CAHAI2020}, and existing and ongoing standardization efforts of both European and international standardization bodies, most notably the European Committee for Standardization (CEN), the European Committee for Electrotechnical Standardization (CENELEC), the International Standardization Organization (ISO), the International Electrotechnical Commission (IEC), and the Institute of Electrical and Electronics Engineers (IEEE)~\cite{IEEE2019}. 

However, we would like to stress once more that transparency may also open up avenues for exploitation  and is hence not always desirable.  For instance, transparency with regard to the algorithm used changes our game-theoretic proposal. It is relatively clear that deterministic classifiers that act as agents in pure-strategy games are  gamed if the strategies are known---with particular ease if the game is repeated. Take the example of AV owners inviting others for a ride to increase the value of human welfare on board  and bias the odds in favor of passengers and against pedestrians.  But transparency can also pose an issue with randomized classifiers in mixed-strategy game settings. Our discussion assumed that the artificial agent cannot infer any information on the merit to  force a deterministic behavior. However, given easy access to big data, it is conceivable that \ML\/ techniques could produce models  to predict the merit---or negative merit, i.e., fault---of the human agents involved. For instance, provided sufficient training data, individuals may use \ML\/ models to infer and discover  which attributes an AI system regards as leading to meritorious outcomes in a certain situation  and adjust their strategy accordingly.  In particular, knowledge that the system is applying logistic-regression is sufficient to use many of  the methods in the literature to find the most influential attributes,  allowing people to artificially boost their values for those.  Even if the model used by the AI is unknown, the use of logistic-regression,  feature selection, or methods based on information gain can reveal to human agents the  most influential attributes used by the AI to ascertain  their merit, leaving plenty of room for human manipulation. Another question is whether, in settings where merit is not ascertainable, regulators should require AI designers to install the best predictive models of human fault in their systems.  This would encourage continuous efforts to at least approximate non-ascertainable attributes  with a view to arrive at more precise decisions. Ultimately, these are questions for policymakers to decide,  we just want to raise awareness to potential problems stemming from ever-increasing transparency  requirements and suggest that there may be certain trade-offs to consider.

Third, although we have established that randomized decisions are mathematically optimal and probably acceptable,  we face a substantial implementation problem:
According to the Church–Turing thesis, no function computable by a finite algorithm  can implement a true random oracle. In other words, true randomness in the mathematical sense is impossible to achieve in a deterministic  Turing machine or von Neumann architecture. The closest thing to generating random sequences are cryptographic hash functions.  These are merely pseudo-random sequences,  which are still susceptible to attacks and may become truly vulnerable with certain advances in quantum  cryptography~\cite{Goldreich1988,Buchmann2017}. Random oracles have been used to prove the security of some cryptographic protocols formally,  but these proofs have come under severe criticism~\cite{Canetti2004} and become subject of heated debates in the cryptographic community~\cite{Koblitz2015}. Despite their theoretic weakness, most users of cryptographic protocols remain unconcerned that,  in practice, random oracles are replaced by cryptographic hash functions~\cite{Koblitz2015}.  

A promising emerging solution to the problem is the novel trend of stepping away from the deterministic  Turing machine and the von Neumann model towards the use of hardware random number generators. These use random numbers from a physical process---such as thermal noise---to harvest entropy~\cite{Goichon}.  Although they still do not qualify as mathematically random,  these sequences have in some cases been extremely unpredictable---but note that from a cryptographic perspective,  they have turned out to be less secure than originally anticipated~\cite{Kim2013}.  In any case, harvesting entropy may also lead to problems:  for example, large logs may be required to satisfy the trustworthy AI requirements,  which designers can review to fix a problem in the event of tragedies.

There are some deterministic processes, like cellular automata, which can be used to generate unpredictable sequences~\cite{Wolfram}. But still, how current computers would generate true random sequences raises a series of philosophical questions: If hardware random generators in combination with pseudo-random generators replaced the notion of pure random oracle for lack of a better implementation in our current hardware, and machines proliferated around us making decisions about our lives, the world could become potentially distressing.  In our previous AV example, perhaps owners would be particularly interested in breaking into the code of the pseudo-random number generator in their car to anticipate its decisions---even with entropy sources attacks are possible~\cite{Kim2013}. Similarly, law firms could invest resources in finding the pseudo-random code behind AI judges to find the best strategies for their clients. In such cases, unpredictable AI rulings would be preferable. Otherwise, too much transparency of the model behind the AI may result in a learnable behavior, from which \ML\/ algorithms could build an accurate model. The model could be used to take advantage and be able to present the attributes of a case much more favorably than otherwise. Note that randomization is used to combat manipulative strategies even in the context of human judges: Such is the perception that knowledge of the judge traits can bias the trial, that human judges for federal cases are typically assigned randomly in the US.

\section{Empirical study}
\label{casestudy}

There is an increase of research on moral human-robot interaction~\cite[and references]{MalleScheutzArnold} where variants of the trolley dilemma~\cite{Foot1967,SW2020} are used to elicit humans' perceptions. For instance, it has been consistently found that humans expect robots to make utilitarian choices~\cite{Bonnefon,Estivill-CastroEthics,MalleScheutzArnold}.

However, we designed our questionnaire to gather empirical evidence on the level of trust society places in technology---or distrusts human decision-makers acting without technological support. We were interested in situations where AI, artificial agents, embedded systems,  and robotics could substitute human agents. Some questions have been motivated by examples provided by reports on the use of AI tools in assisting judges in sentencing and bail decisions~\cite{Kugler2018}. While we expected most people to agree that, in principle, machine assistance improves human decision-making performance, we suspected differences in opinion as to the extent such technologies should substitute human decision-makers, particularly when decisions directly impact another person. To measure such differences, we selected three different groups of participants, each with different levels of exposure to the debate on moral decision-making by AI systems. While our initial questions assumed deterministic AI behavior, later questions touched on the issue of randomness.

\subsection{Method}
Prior to the survey, we familiarized participants with the EOP to provide a relatively simple, underlying framework to avoid confrontation with the many fairness definitions against which classifiers' performance may be judged~\cite{Heidari2019}. For the introductory paragraph and the questionnaire, see Appendix~\ref{ap_research_inst}. 

We recruited  participants through four different methods. The first set of participants were Australian full-time employed high-school teachers holding a university degree, who participated in a series of professional development workshops to include or expand the use of technology in their pedagogy. They came from 34 different schools with their levels of delivery including junior secondary, year~7 to~9, year~7 to~12, year~10, STEM~P-6, year~9 to~12, VET, primary school, year~8 to~12, and K~6.  Participation in the survey was an optional activity offered in the workshops. Out of 65 invited participants, 53 replied to our questionnaire and 12 declined. We will refer to this group of participants as \emph{teachers} in Section~\ref{Analysis}.

The second means of recruitment was an online survey ~\cite{MalleScheutzArnold} using  SurveyMonkey Inc.~\footnote{The data  was collected using SurveyMonkey~Audience. We chose Age Balancing: Basic, Gender Balancing: Census,
COUNTRY: United States (USA) - SurveyMonkey, HOUSEHOLD INCOME: $\$50k-\$200k+$,
AGE: $25-100+$, Education: Graduate degree.  Information on how respondents are recruited to SurveyMonkey is available here:
www.surveymonkey.com/mp/audience}, through which we collected $124$ responses.
We limited the group of acceptable respondents, since we believe the concepts and issues regarding the potential applications of AI in supporting and replacing human decision-making are not trivial, particularly regarding moral standing and fairness: (1) One excluded respondent class were children, as their views on robots significantly differ from adults~\cite{BartlettECS03,KahnFH03}. (2) We selected respondents from the US, who were above the age of 25, held a graduate degree, and had a household income of \$50K-\$200k. Our choice of participant group was motivated by the fact that individuals' engagement in policymaking largely depends on their socialisation and education, and we targeted the US because of its culture of deep and genuine citizen participation~\cite{AustralianParliMentReport}.
Many reports~\cite{AndersonFACTTANK} highlight that there is different levels of adoption of technology in the US, but we assume that the minimum household income and their use of  SurveyMonkey represents a homogeneous group in this regard. We note that other on-line experiments have a similar sample size and data collection approach for a generic view by humans~\cite{MalleScheutzArnold}. We will refer to this group of participants as \emph{online respondents} in Section~\ref{Analysis}.

Third, we performed a robotics demonstration on Nathan Campus, Brisbane, at University Open Day. Our robots engaged visitors in conversation by randomly picking shorter versions of the survey questions. 17 visitors became curious and agreed to complete the survey. We will refer to this group of participants as \emph{Open Day visitors} in Section~\ref{Analysis}.

Finally,  we collected 13 surveys from IT academics, whom we invited to our department's research seminar. We will refer to this group of participants as \emph{academics} in Section~\ref{Analysis}.

All survey participants were provided with a consent and information
sheet as per the conditions of our Research Ethics Committee's approval.
Thus, participation was completely voluntary, anonymous, and could stop at any time. For the SurveyMonkey collection, we ensured respondents could not go back and alter their answers for previous questions. The other data collection methods used printed versions of the questionnaire, so we cannot guarantee people did not revise their answers. However, we feel this is very unlikely as we found no amendments when capturing the answers.

Most of the questions in our questionnaire can be considered Likert-type scale questions (i.e., we give the survey participants five worded points they can choose from). Whenever possible, we will avoid debate about whether the data collected should be considered as interval-level data or ordered-categorical data. As mentioned above, we will use a five item option with a clear middle neutral point. Questions are symmetrical although potentially not equidistant; their presentation aimed for respondents to assume equidistance. We will use  diverging stacked bar charts when presenting the descriptive statistic analysis
of ordinal data~\cite{JSSv057i05} and bar charts in the remaining cases. Since we will be comparing different groups of respondents for the same question, 
the responses can not be paired (sample sizes are different), and we are not making any assumptions about the normality of the data. We have overlapping samples which we will analyse for statistical difference of medians using the
non-parametric Mann-Whitney U test\footnote{We replicated the calculation using the Mann-Whitney U Test Calculator from https://www.socscistatistics.com
and R~\cite{citingR}.}~\cite{DerrickWhiteLikert}.

Given two random variables $X$ and $Y$ with their cumulative distribution functions $f$ and $g$, respectively. For ordered samples $x_1, \ldots , x_n$ and $y_1, \ldots , y_m$ of $X$ and $Y$, respectively, let the statistic $U$ count the number of how many times a $y$ value precedes an $x$ value. The Mann-Whitney U Test uses the statistic $U$ for testing the hypothesis $f=g$~\cite{mann-whitney}. The null hypothesis $H_0$ in our tests is going to be that the two cumulative distribution functions are the same (i.e., two different groups have the same scoring tendency). In all of our tests we will use the significance level $\alpha=0.05$. Furthermore, we will specify the \emph{$p$-value}, the probability (assuming that $H_0$ is true) that we obtain a
test statistic that is at least as extreme as the observed value of the test statistic. If the $p$-value is less than the level of significance, we reject the null hypothesis and accept the alternative hypothesis.

\subsection{Analysis}
\label{Analysis}

\subsubsection{Evidence for preference for automated judgments}

\begin{figure}
\centering
\subfloat[\label{Question1divergent}Question~\ref{Q1}: Who is fairer, humans or computers, to implement EOP.]{
\includegraphics[width=0.45\textwidth]{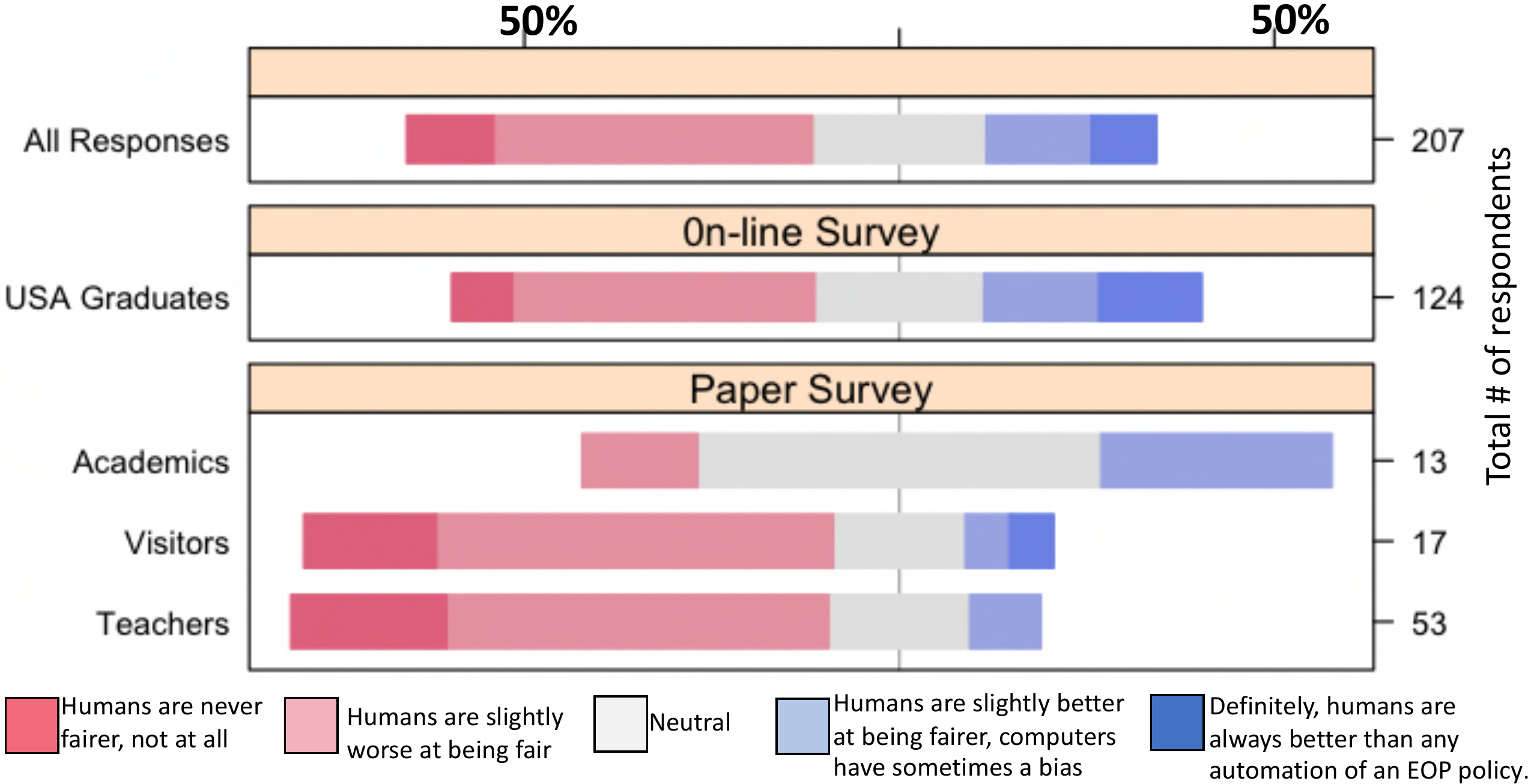}} $\;\;$
\subfloat[\label{Question03divergent}Question~\ref{Q3}: How to select which passenger should be left behind from a flight.]{
\includegraphics[width=0.45\textwidth]{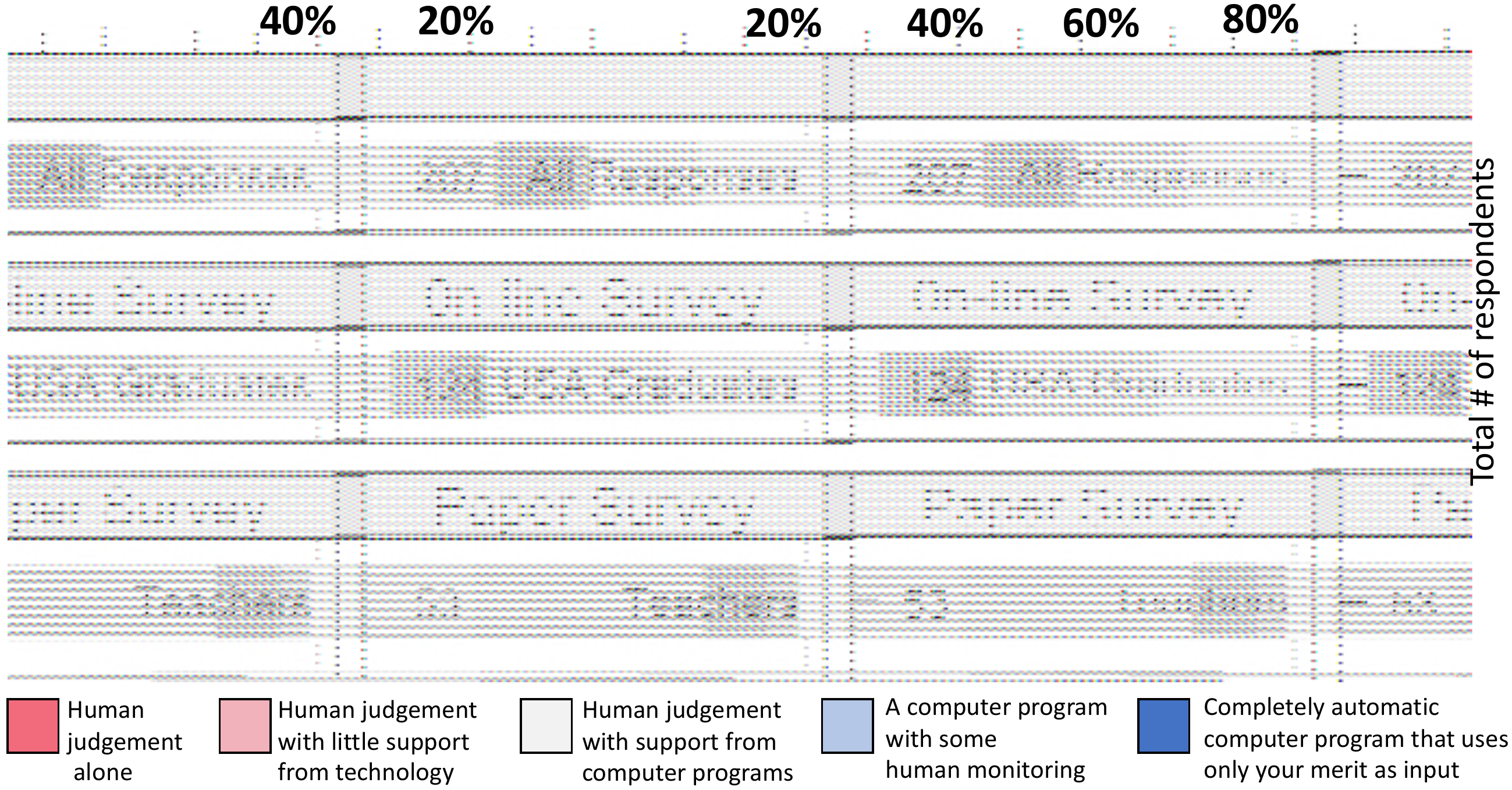}}
\caption{\vspace{-1mm}\label{Q1ANDQ3divergent} Divergent bar charts of responses to Question~\ref{Q1} and Question~\ref{Q3}.
}
\vspace{-2.5mm}
\end{figure}

The results in this subsection show that some groups of participants' distrust in humans leads them to prefer AI over human decision-makers.
Figure~\ref{Question1divergent} shows the divergent bar chart of responses
to Question~\ref{Q1} from our groups  as percentages.
Eleven teachers felt humans are never fairer, and none felt humans are better than automation. A small percentage ($13.7\%$) of SurveyMonkey respondents were of the opinion that humans can be fairer, but $40.3\%$ believed humans are slightly worse at being fair. Teachers' and Open Day visitors' median and mode indicated that humans  were somewhat worse at being fair, while the academic view was neutral. The difference between teachers and academics is statistically significant ($U=137$, $p=0.00042$) and so is the difference between Open Day visitors and academics ($U=49.5$, $p=0.0057$).

\textbf{Discussion:}
Figure~\ref{Question1divergent} 
shows results for Question~\ref{Q1} and  assesses individuals' level of trust in human decision-makers not supported by technology. The responses accumulate away from the use of human judgment alone. In this case, the results are consistent with the generally accepted view~\cite{Kugler2018} that machines analyze facts without human bias, irrationality, or mistakes. Thus, it seems
teachers and Open Day visitors perceive machines as more impartial than humans.
Although this view  may shift due to the growing recognition of the need for a human-in-the-loop approach~\cite{EUReport}, these groups of respondents do not seem to be dissuaded from automation.

Question~\ref{Q3} (refer to Figure~\ref{Question03divergent}), which concerns the automation of decisions on regrouping passengers across flights, also shows preference for AI-decisions over human involvement. Teachers' median and mode are again for a computer program with some human monitoring. Not a single academic selected only human judgment. Open Day visitors' median and mode are also for this level of automation, and so is the mode of the online survey.
Teachers' preference for automation is significantly different 
from that of online respondents ($U=1921$, $p=0.0001$) and Open Day visitors ($U=302.5$, $p=0.04338$).

\begin{figure}
\centering
\subfloat[\label{Question11divergent}Question~\ref{Q11}: Who should evaluate a travel visa application, human or artificial agent.]{
\includegraphics[width=0.45\textwidth]{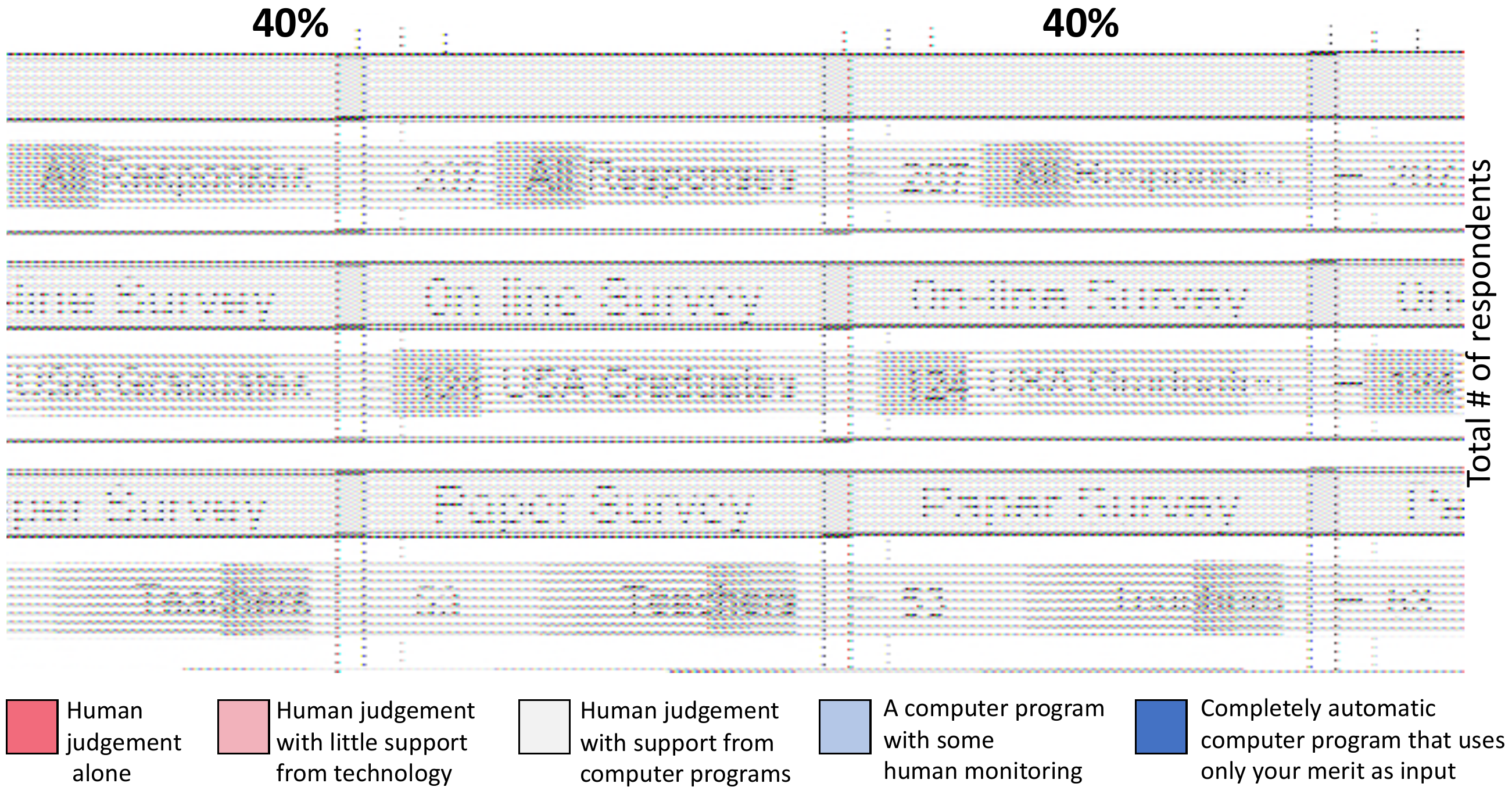}} $\;\;$
\subfloat[\label{Question12divergent}Question~\ref{Q12}: Who should decide over awarding  a credit-card and its limit, humans or machines.]{
\includegraphics[width=0.45\textwidth]{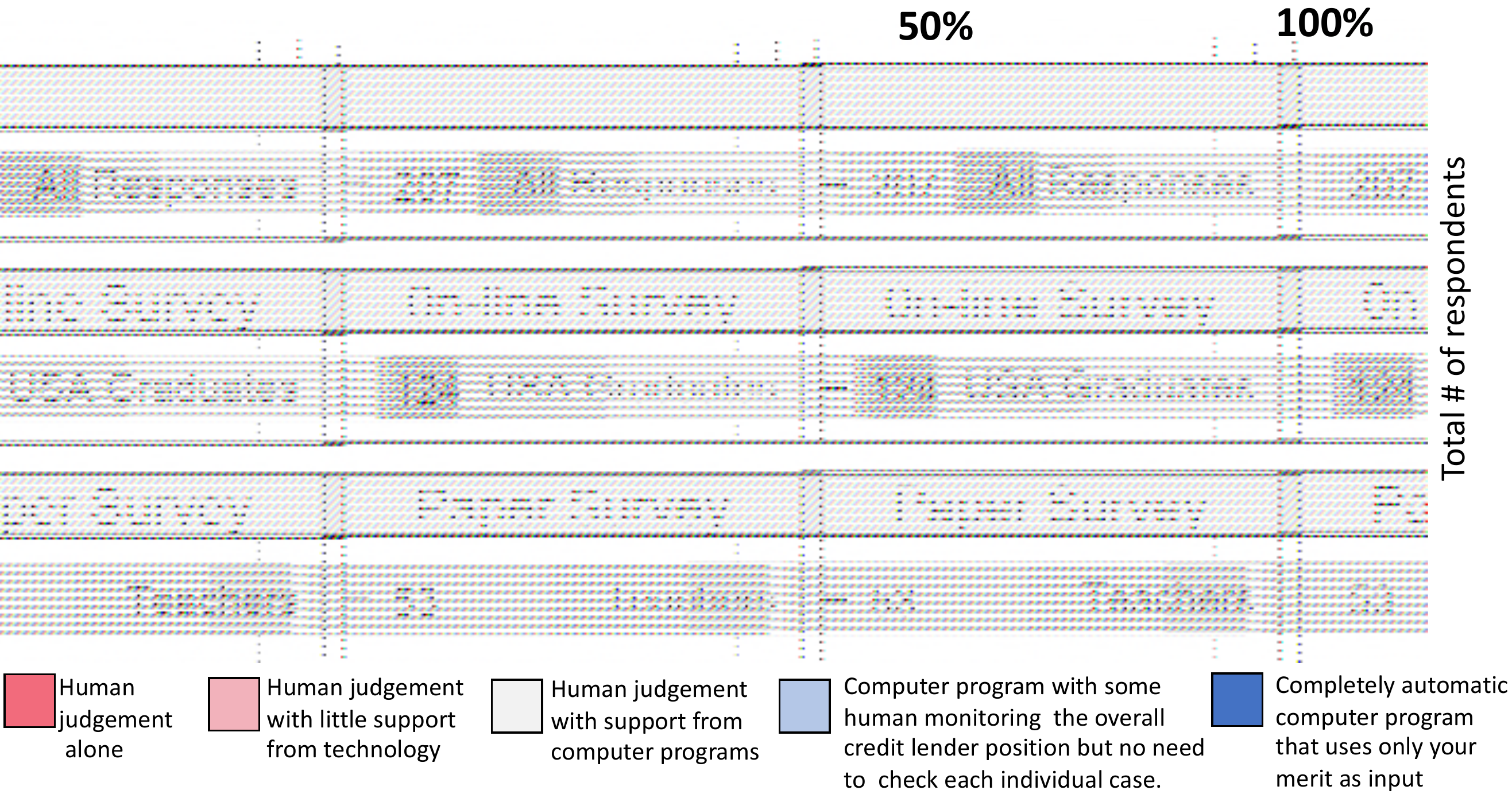}}
\caption{\vspace{-1mm}\label{Q11ANDQ12divergent} Divergent bar charts of responses to Question~\ref{Q11} and Question~\ref{Q12}.
}
\vspace{-2.5mm}
\end{figure}

The responses for Question~\ref{Q11} (see Figure~\ref{Question11divergent})  on automating visa applications show an almost identical pattern to Question~\ref{Q3}.  Responses are clearly in favor of automation and reject human judgment: Teachers' and Open Day visitors' median and mode unambiguously reflect this preference, and the mode of online respondents is also for automation (the median is neutral).  To a slightly lesser extent, academics also favor automation. Both Question~\ref{Q3} and Question~\ref{Q11} received a non-trivial number of responses on the extreme values, but mainly advocating fully automated decisions. Albeit in the opposite direction, we will see below the results for Question~\ref{Q2}, which generally favors human judgment. But even in    Question~\ref{Q2}, there are also non-trivial numbers of those who would prefer the selection to be entirely without human intervention. The difference between the responses of teachers and online respondents is again
statistically significant ($U=1968$, $p=0.0001$). But interestingly, here, Open Day visitors' high preference for an AI-judgment is statistically significant with respect to online respondents ($U=696$, $p=0.02382$).

Question~\ref{Q12}---concerning decisions on credit-card applications---shows responses even more in favor of  automation (refer to Figure~\ref{Question12divergent}). The median and mode for teachers, academics, and Open Day visitors indicates ruling over credit-card applications should be a matter for computer programs with some human monitoring, yet without checking each individual case. Almost all teachers favor automation, clearly rejecting human judgment even if supported by technology. Online respondents, academics, and Open Day visitors also show a strong tendency towards automation: human judgment is less than $8\%$ of the responses for all participants. Even given all groups' preference for the use of technology, note that teachers' are in favor of automation to such an extent that the difference between their views and that of online respondents is statistically significant ($U=1967.5$, $p=0.00001$).

\textbf{Discussion:}
A possible explanation for such a high level of acceptance of automated decision-making over human judgment in this scenario is that the massive benefits of automation in the credit-card industry pushed an early penetration of  \ML\/ technology in the sector: \ML\/ has been applied to credit-card application assessments for more than 25 years~\cite{Carter87,Davis92}.
In this setting,  there have been calls for accepting technology with little questioning~\cite{Rosenberg2008}. 

\begin{figure}
\centering
\subfloat[\label{Question7divergent}Question~\ref{Q7}: How much automation when ruling over child adoption.]{
\includegraphics[width=0.45\textwidth]{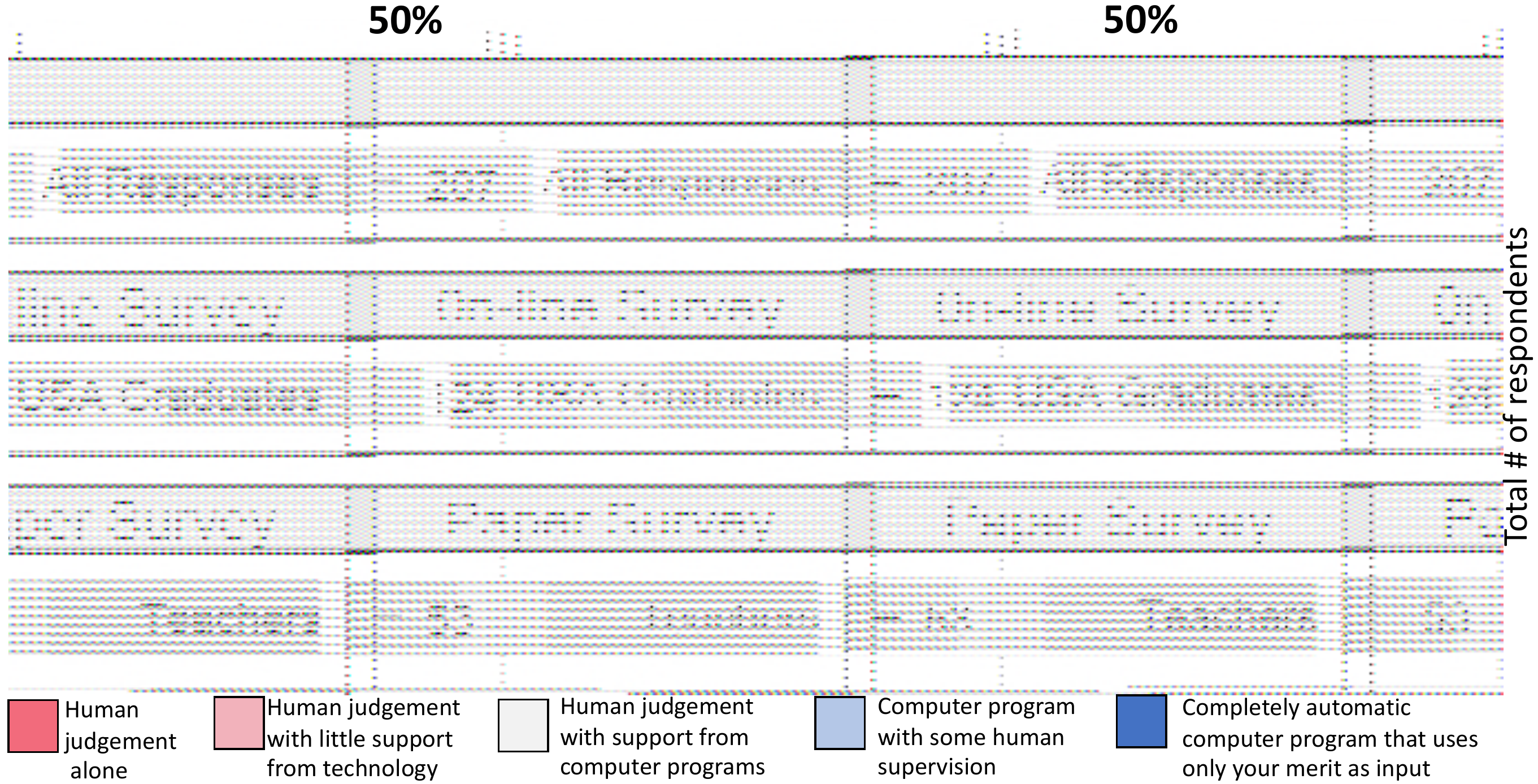}} $\;\;$
\subfloat[\label{Question8divergent}Question~\ref{Q8}: How much automation when judging over a bail hearing.]{
\includegraphics[width=0.45\textwidth]{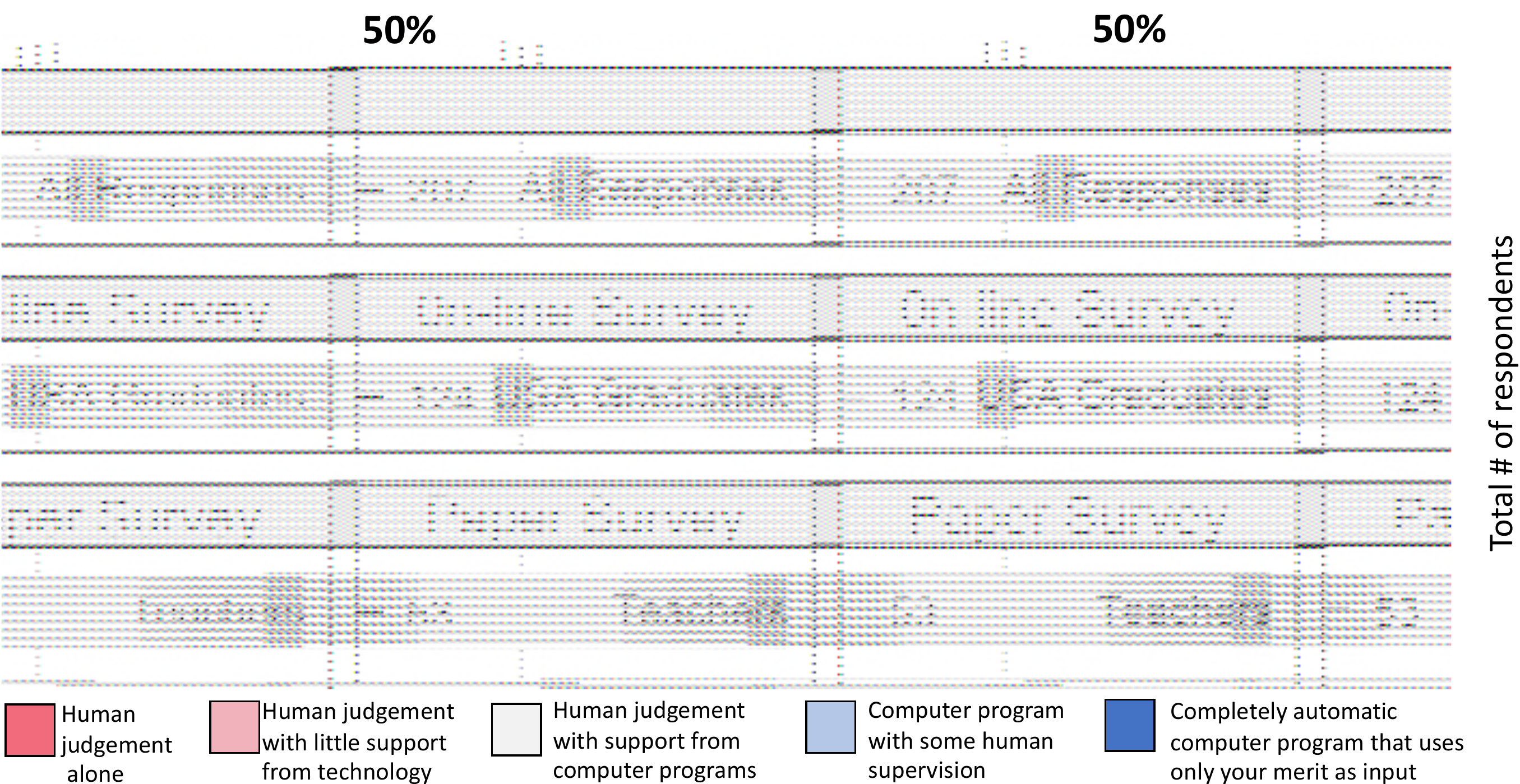}}
\caption{\vspace{-1mm}\label{Q7ANDQ8divergent} Divergent bar charts of responses to Question~\ref{Q7} and Question~\ref{Q8}.
}
\vspace{-2.5mm}
\end{figure}

With less overall preference for a machine-driven decision, Figure~\ref{Q7ANDQ8divergent} shows the results of Question~\ref{Q7} and Question~\ref{Q8}.
Figure~\ref{Question7divergent} and Figure~\ref{Question8divergent} show
almost identical proportions and the positions of the bars. For online respondents and for Open Day visitors, there is strong centrality, and a firm rejection for complete automation of the decision-making. Less than $5\%$ of online respondents chose the fully automated option for both of these questions. However, the teacher group shows a strong preference for the decision to be strongly guided by full automation or only some human supervision. Responses for teachers on both questions have a median and mode for using a computer program with some human supervision for both scenarios.
Teachers' preference for AI-judgments is statistically significant with respect to online respondents ($U=1360.5$, $p=0.00001$ for Question~\ref{Q7} and
$U=2184$, $p=0.00004$ for Question~\ref{Q8}) and with respect to Open Day visitors for Question~\ref{Q7} ($U=143.5$, $p=0.00001$) but not for  Question~\ref{Q8}.

\begin{figure}
\centering
\subfloat[\label{Question14divergent}Question~\ref{Q14}: Who assesses fairly a funding application from a school, humans or artificial agents.]{
\includegraphics[width=0.45\textwidth]{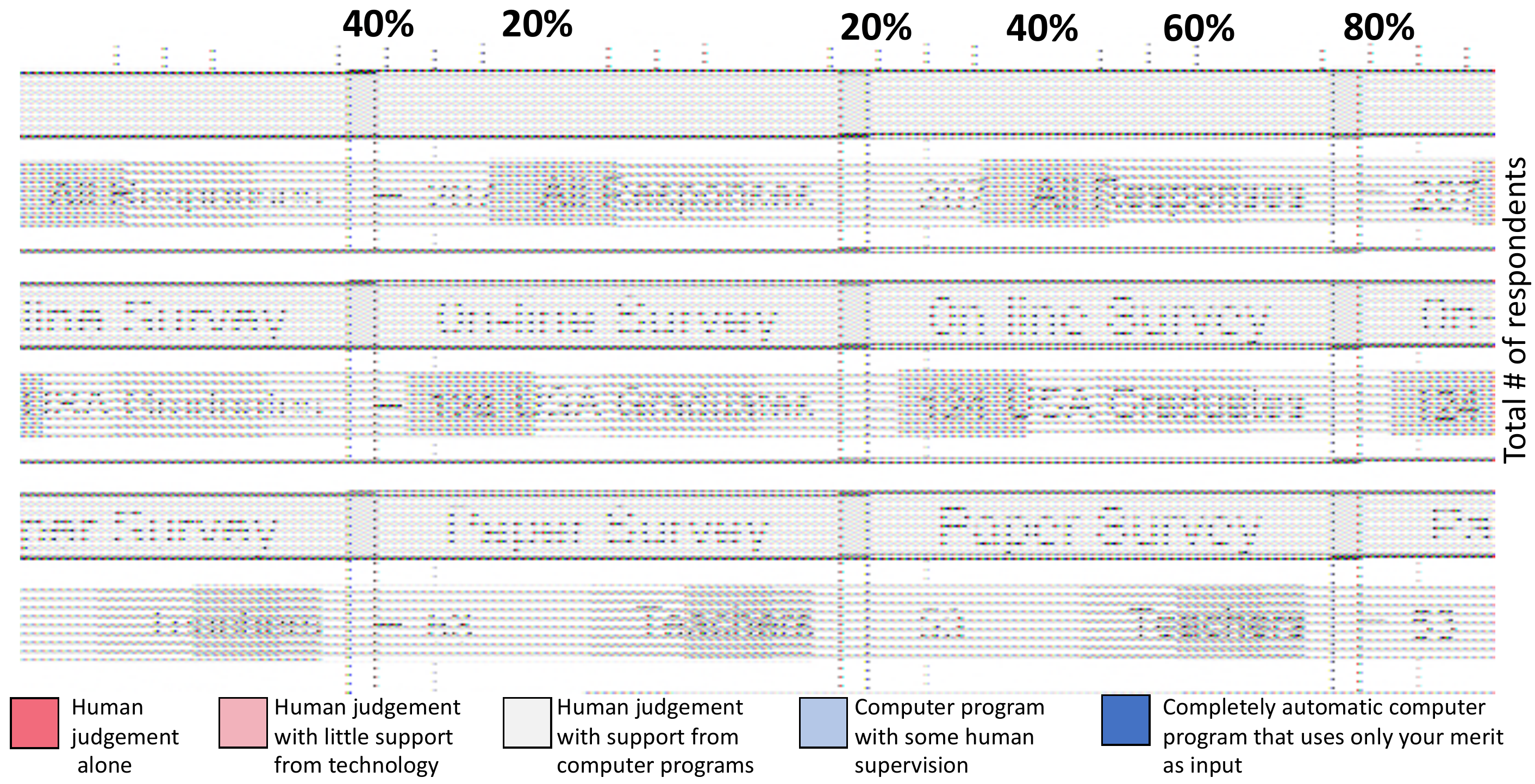}} $\;\;$
\subfloat[\label{Question21divergent}Question~\ref{Q21}: Who is fairer at specifying the sentences for a convicted criminal.]{
\includegraphics[width=0.45\textwidth]{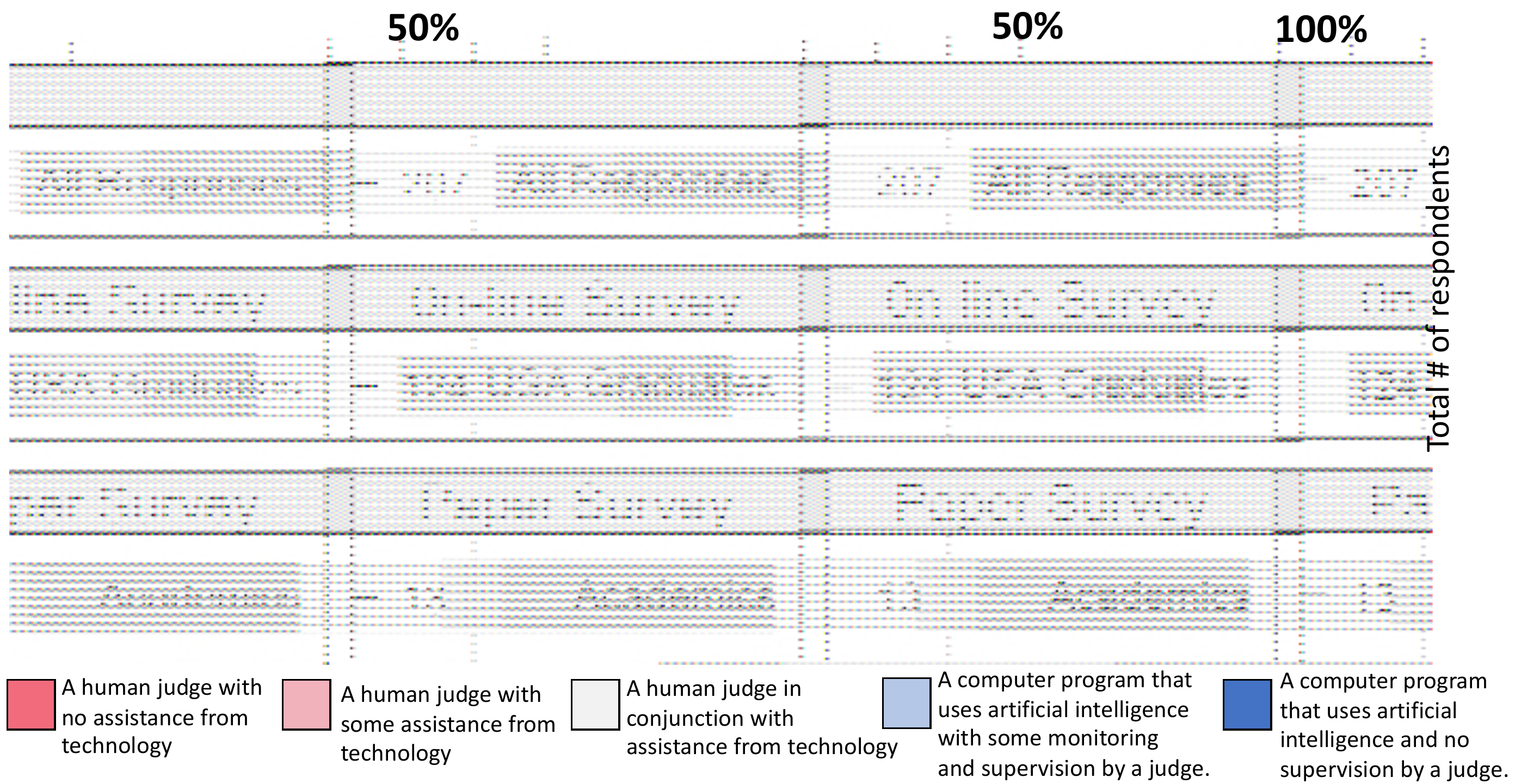}}
\caption{\vspace{-1mm}\label{Q14ANDQ21divergent} Divergent bar charts of responses to Question~\ref{Q14} and Question~\ref{Q21}.
}
\vspace{-2.5mm}
\end{figure}

The results for Question~\ref{Q14} also show central/neutral responses for Open Day visitors and online respondents (see Figure~\ref{Question14divergent}).
However, once more, teachers show a statistically significant higher preference for automation over online respondents ($U=2276.5$, $p=0.00124$)  and Open Day visitors ($U=272.5$, $p=0.0151$).
Question~\ref{Q21} (see Figure~\ref{Question21divergent}) shows the public's tendency to centrality with respect to automating sentencing decisions, 
with a very slight shift in favour of human intervention:  Open Day visitors' median and mode and online respondents' mode indicates a preference for 
technologically assisted human sentencing decisions. On the other hand, the median and mode of academics and teachers is clearly in favor of AI-enabled, 
automated decision-making with some monitoring and supervision by a judge.
Teachers' preference for an AI-judge is statistically significant over that of online respondents ($U=1778$, $p=0.00001$) and Open Day visitors ($U=228$, $p=0.00236$). Academics' support for the AI-judge is even stronger: None of them contemplates a position below neutral. Hence, academics' preference for automated decisions in this setting is statistically significant not only with respect to online respondents ($U=123$, $p=0.00001$), but also Open Day visitors
($U=12$, $p=0.00001$) and teachers ($U=120.5$, $p=0.00016$).

\textbf{Discussion:}
The teachers were surveyed after a professional development day for the inclusion of ICTs in the teaching-learning process. Some of the technologies  demonstrated and discussed were virtual-reality, robotics, and cloud tools. Therefore, they may hold high hopes for the automation of judgment. Alternatively, ICT academics and teachers may be more aware of works on bounded rationality---suggesting that cognitive limitations prevent humans from being fully rational~\cite{Simon1957,Lindblom2017}---and studies indicating that
humans suffer from predictable biases that influence judgment~\cite{KahnemanTversky1979}. These insights presumably encourage them to be more cautious and meticulous when evaluating and designing decision-making methods that may affect human beings. The fact that the systematic design of such methods is typically supported by computer programs may provide a further explanation for this groups' high preference for automation.

\subsubsection{Evidence for preference for human involvement in judgments}
\begin{figure}
\centering
\centering
\subfloat[\label{Question02divergent}Question~\ref{Q2}: Who is fairer at evaluating command of foreign language.]{
\includegraphics[width=0.45\textwidth]{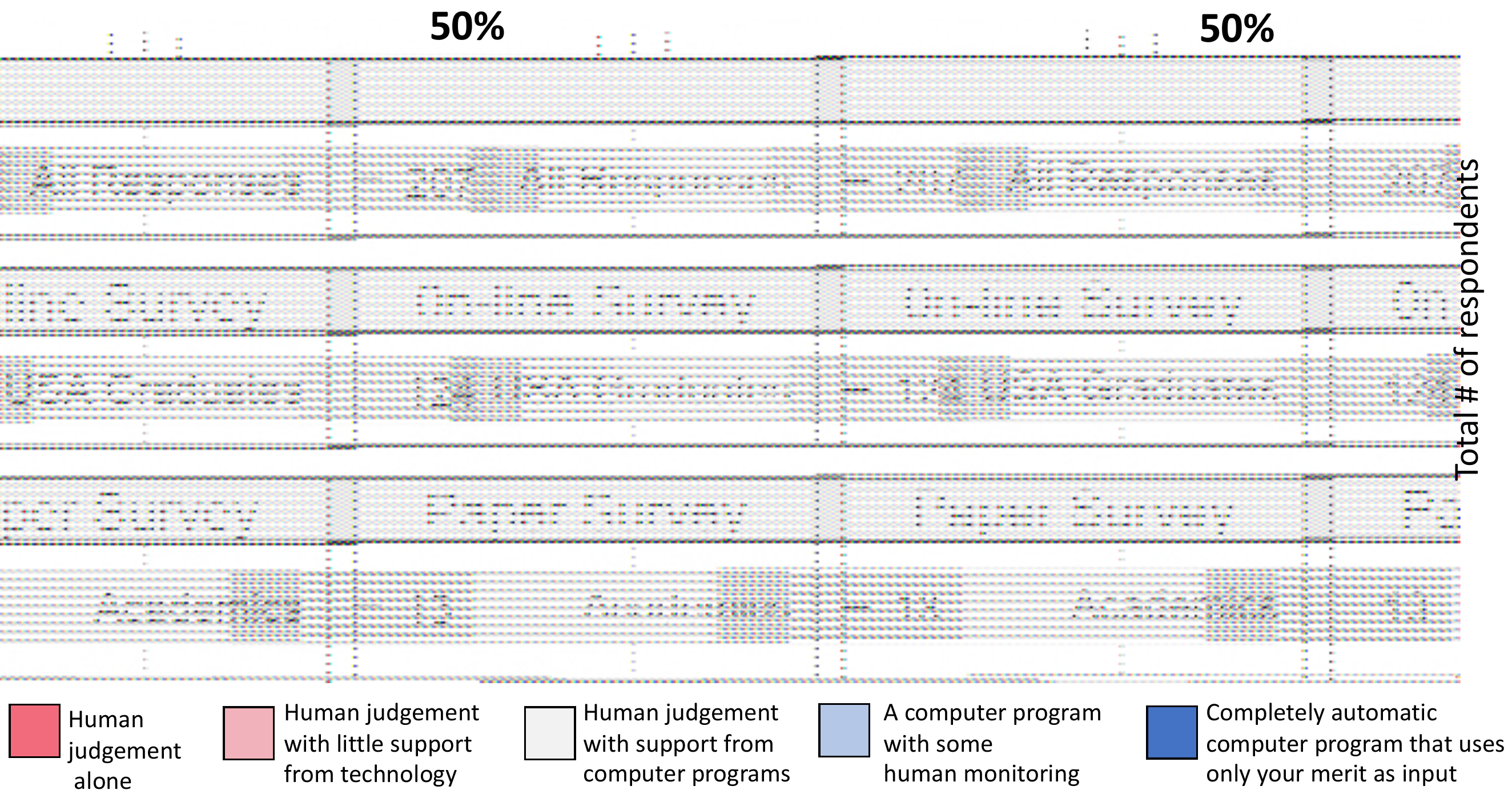}} $\;\;$
\subfloat[\label{Question05divergent}Question~\ref{Q5}: Would a minority person be judged fairly by human/artificial security officer.]{
\includegraphics[width=0.45\textwidth]{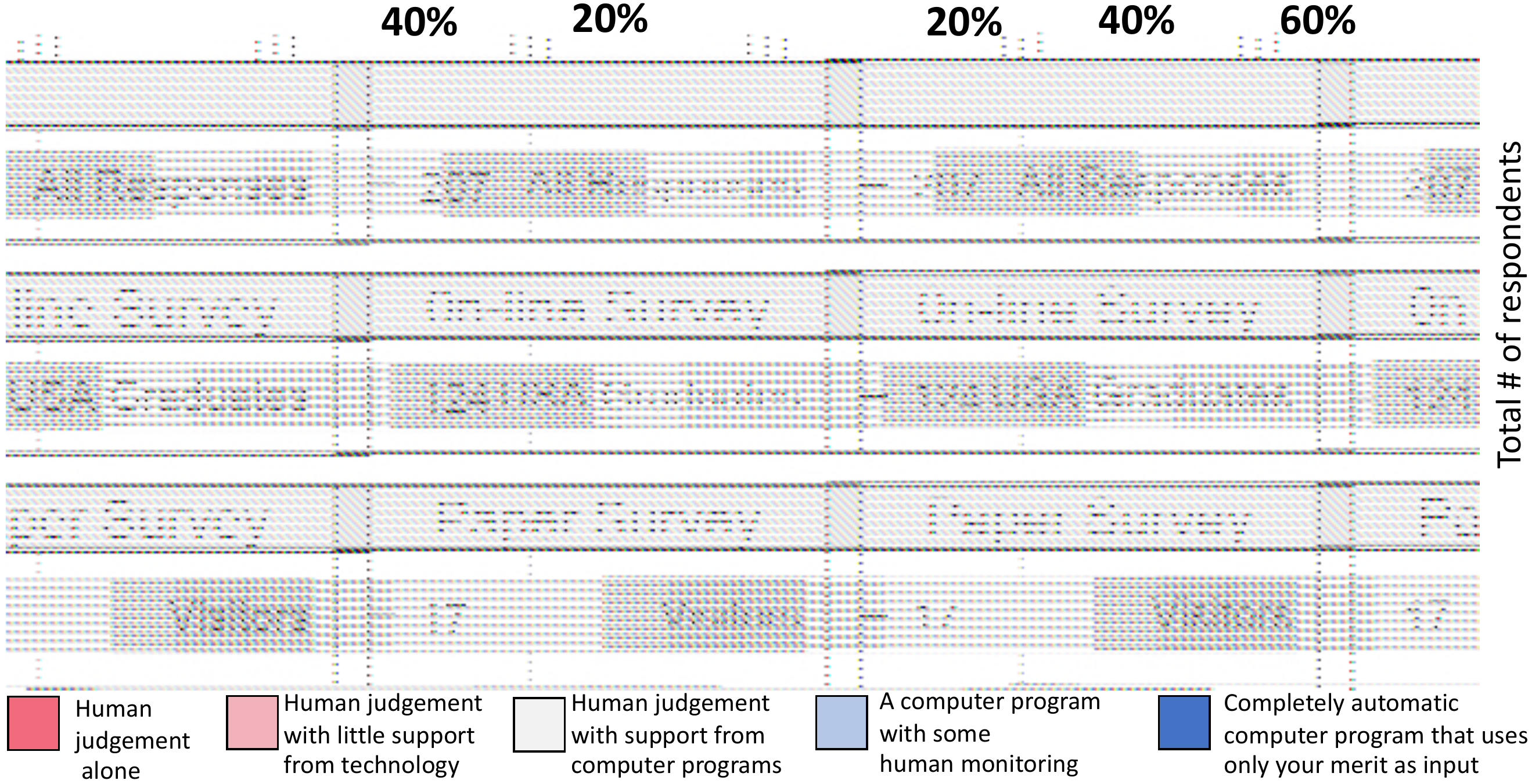}}
\caption{\vspace{-1mm}\label{Q2ANDQ5divergent} Divergent bar charts of responses
to Question~\ref{Q2} and Question~\ref{Q5}.
}
\vspace{-2.5mm}
\end{figure}

Let us now turn to the results showing that some groups of participants exhibit distrust in AI decision-makers and hence prefer their human counterparts. 
Figure~\ref{Question02divergent} shows the corresponding divergent bar chart of responses for Question~\ref{Q2}, which aimed to find out the public's preferred level of automation in the context of a language test. The accumulated results did not swing either way: Teachers preferred some automation, academics even more. Their mode and median were for a computer program with some human monitoring.  Open Day visitors' view was in-line with the online survey. The remarkable point here is the issue at the extremes: In all data collections, there is some non-trivial number of preferences for exclusively human judgment without technological support. Apart from academics, all other respondents rejected total automation.

Although the results we discuss next show a tendency for a neutral response---with some involvement of technology---they also indicate a clear desire by participants to maintain some human involvement, rejecting decisions by fully autonomous AI decision-makers. Question~\ref{Q5} considers a scenario in which a subject, who belongs to a visibly distinct group, needs to request assistance. Some of the responses here show strong evidence that complete reliance on human judgment is unacceptable. The median and mode for both online respondents and Open Day visitors are that if they were a minority, they would prefer the involvement of computer programs with some human supervision.
Teachers' and academics' responses for this question are remarkably neutral (refer to Figure~\ref{Question05divergent}).

\begin{figure}
\centering
\subfloat[\label{Question04divergent}Question~\ref{Q4}: Ruling on discipline measures for a child of a minority group at school.]{
\includegraphics[width=0.45\textwidth]{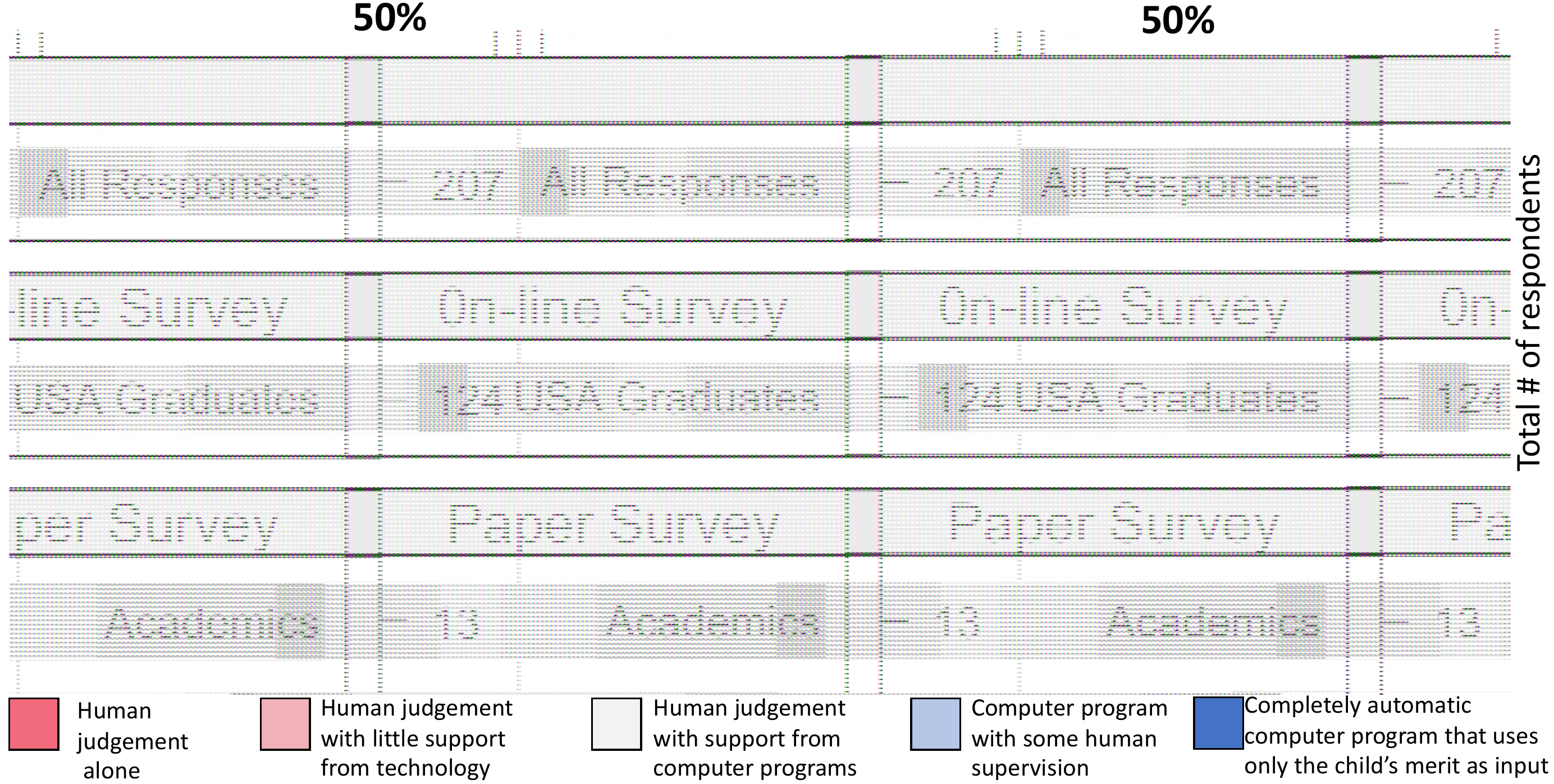}} $\;\;$
\subfloat[\label{Question06divergent}Question~\ref{Q6}: Judging behaviour at nightclub by human/artificial security guard.]{
\includegraphics[width=0.45\textwidth]{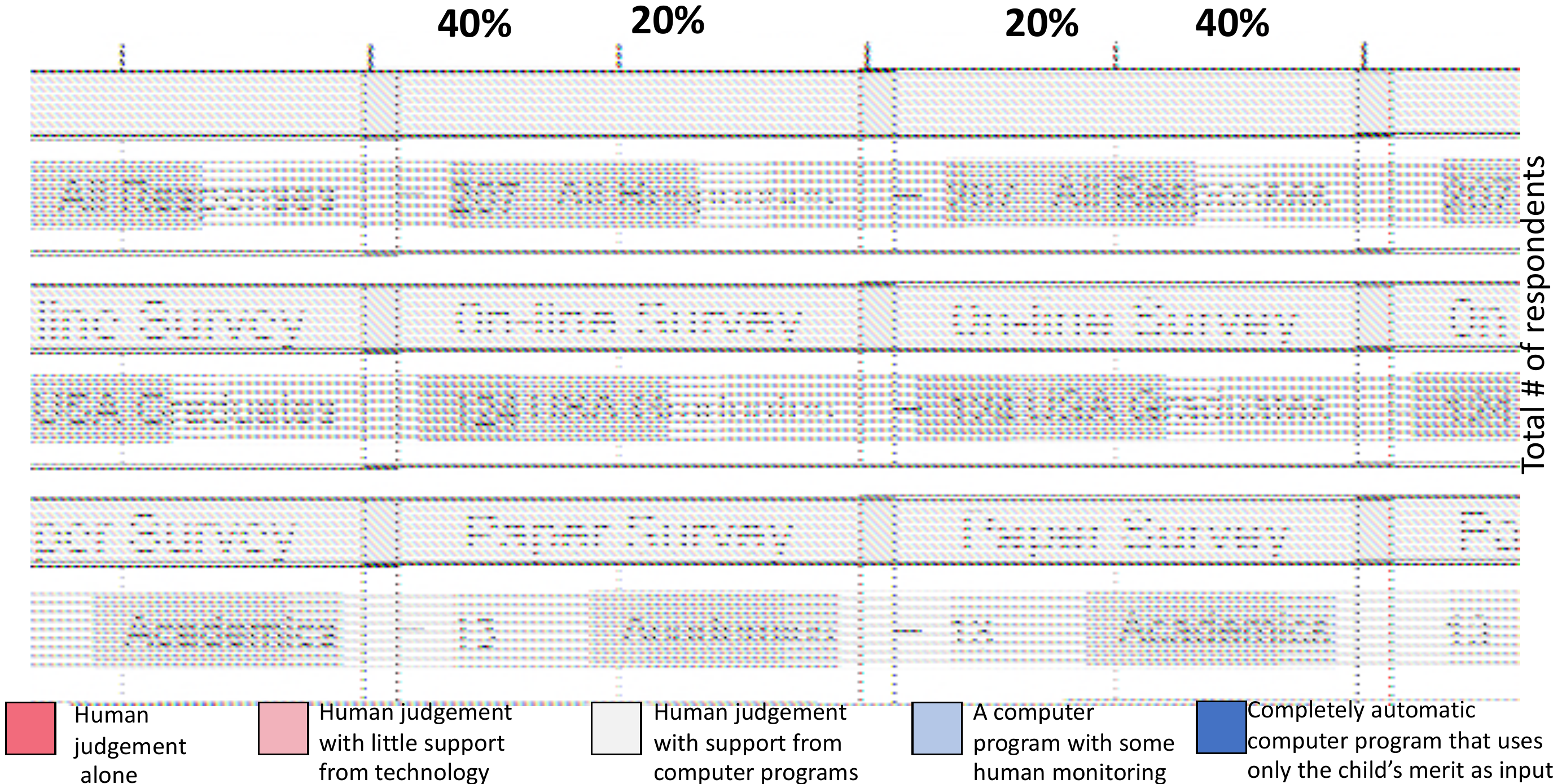}}
\caption{\vspace{-1mm}\label{Q4ANDQ6divergent} Divergent bar charts of responses
to Question~\ref{Q4} and Question~\ref{Q6}.
}
\vspace{-2.5mm}
\end{figure}

Figure~\ref{Q4ANDQ6divergent} shows a global neutral response for two questions.
For Question~\ref{Q4}, most respondents stay away from fully-automated rulings but also dislike purely humans verdicts that lack any support from computer systems. No single respondent from the Open Day visitors elected the fully automatic option. In the US online survey, the utterly automatic option has less than $6\%$ of the respondents. While the median and mode for the online respondents and the Open Day visitors is the neutral option, teachers' responses again show distrust in human decisions and their median and mode are for the use of a computer program with some human supervision for ruling on punishments at school for children (see Figure~\ref{Question04divergent}).
Teachers' preference for technology is statistically significant both with respect to online respondents ($U= 1933 $, $p=0.00001$) and Open Day visitors
($U= 234.5$, $p=0.00159$). Academics are even more in favor of higher levels of machine-based decisions in this setting, and their position is statistically significant both with respect to online respondents ($U= 394$, $p=0.00126$)
and Open Day visitors ($U= 47.5$, $p=0.0044$). A similar picture of neutrality
emerges from Question~\ref{Q6}  (refer to Figure~\ref{Question06divergent}). The responses are more or less even, except again that very few respondents would accept a completely artificial security guard. None of the responses from Open Day visitors corresponds to this option. However, academics remain inclined for automation---none of them selected a human security guard alone. And interestingly, in this question, teachers have shifted to a more even and spread opinion. Nevertheless, the Mann-Whitney U Test does not reveal a statistical difference at the $0.05$ level for any two groups of responses.

\subsubsection{Data on acceptance of technological assistance}
We now show results for questions we later use for contrasting sub-groups between the respondents.

\begin{figure}
\centering
\subfloat[\label{Question09}Question~\ref{Q9}: Discussing your medical condition to a human versus a robot.]{
\includegraphics[width=0.45\textwidth]{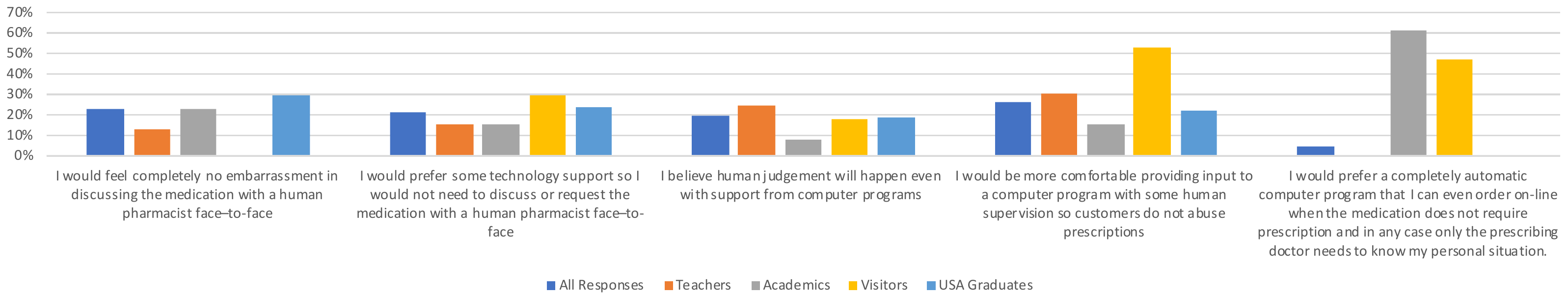}} $\;\;$
\subfloat[\label{Question17}Question~\ref{Q17}:  Has artificial intelligence surpassed human performance for the game of chess.]{
\includegraphics[width=0.45\textwidth]{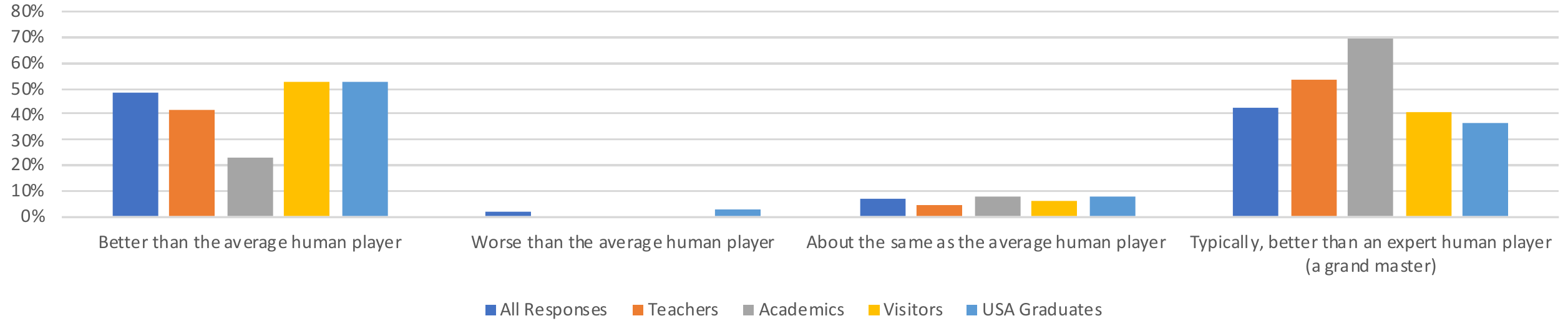}}
\caption{\vspace{-1mm}\label{Q9ANDQ17bar} Bar charts of responses
to Question~\ref{Q9} and Question~\ref{Q17}.
}
\vspace{-2.5mm}
\end{figure}

Question~\ref{Q9} (refer to Figure~\ref{Question09}) shows mixed results.
Although one side indicates complete comfort with disclosing medical conditions to a human pharmacist, while the other an absolute preference for no human involvement in fulfilling a prescription, the options in this question may not be considered symmetric or an ordered categorical scale. Figure~\ref{Question09} shows bar charts of percentages of responses and
we note that
there is a similar dispersion of preferences between teachers' and online respondents' responses. These two groups are comfortable discussing prescriptions with pharmacists. However, Open Day visitors never picked the option to discuss face-to-face with a human and have far more preference for full automation. The mode for academics is also full automation.

\textbf{Discussion:}
We formulated Question~\ref{Q9} as a result of the reports~\cite{Reese2017} around the virtues of using a robot as a point of sale in retail (the first $1,000$ Pepper robots were hired for Nescafe in December 2015 to help customers searching for coffee machines in their appliance stores). This question is also motivated by the suggestion that technological interventions~\cite{Scoglio2019,Farreretal2013} in which humans interact with a robot enable humans to feel more at ease. However, the reverse is also possible (recall the uncanny valley debate~\cite{PIWEK2014271}). Upon reflection, there could have been more questions and research to explore the perception by a human patient to receive fair or appropriate treatment from a robotic nurse versus a human nurse. Biases in medical treatment because of factors that could be considered discrimination are well documented~\cite{Hall2105}.

Question~\ref{Q10} and Question~\ref{Q15}  attempt to measure a construct by which technological assistance to humans is accepted because it has prevailed for several generations. Note that acceptance of new information technologies has been researched extensively~\cite{AgarwalPrasad} but familiarity with the notion of autopilot in planes is not comparable to  familiarity with robots/machines as judges or moral decision makers.  That is, information technologies are appearing at a much faster rate than some innovations in automation. The origins of the plane's autopilot is attributed to the Sperry family (or Corporation) in 1912, and in 1947 a  US Air Force C-53  completed a transatlantic flight apparently with all phases in autopilot. Similarly, cruise control patents appeared in the 50s and 60s, although widespread use of cruise control appeared in the 1070s because of its positive impact on fuel consumption. We speculate that our respondents are all comfortable with the idea that it is advantageous to have autopilots in long flights and that cruise control is a positive option.
\begin{figure}
\centering
\subfloat[\label{Question10}Question~\ref{Q10}: Is the autopilot preferable than not having one.]{
\includegraphics[width=0.45\textwidth]{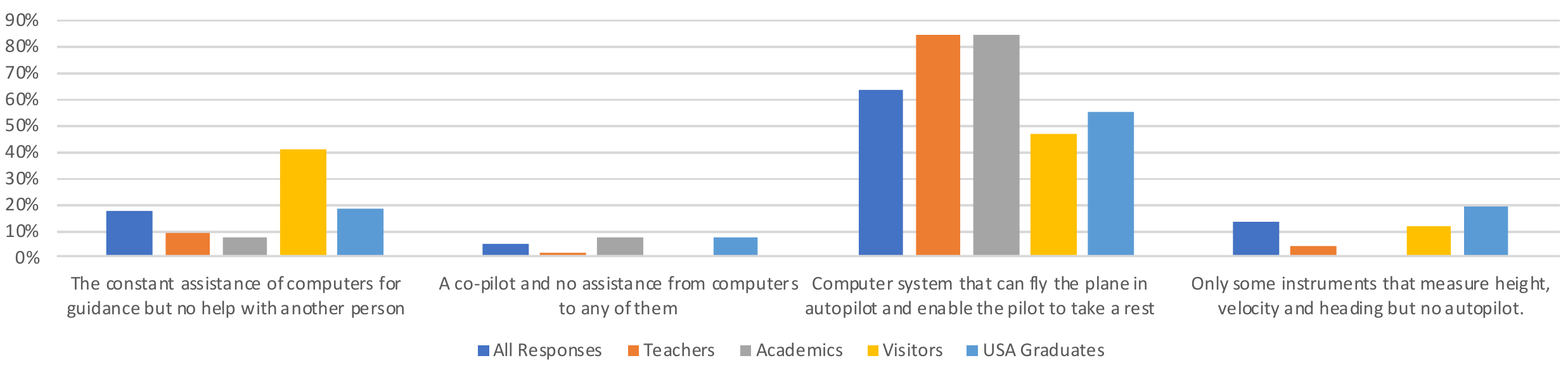}} $\;\;$
\subfloat[\label{Question15}Question~\ref{Q15}: Does cruise control facilitate driving.]{
\includegraphics[width=0.45\textwidth]{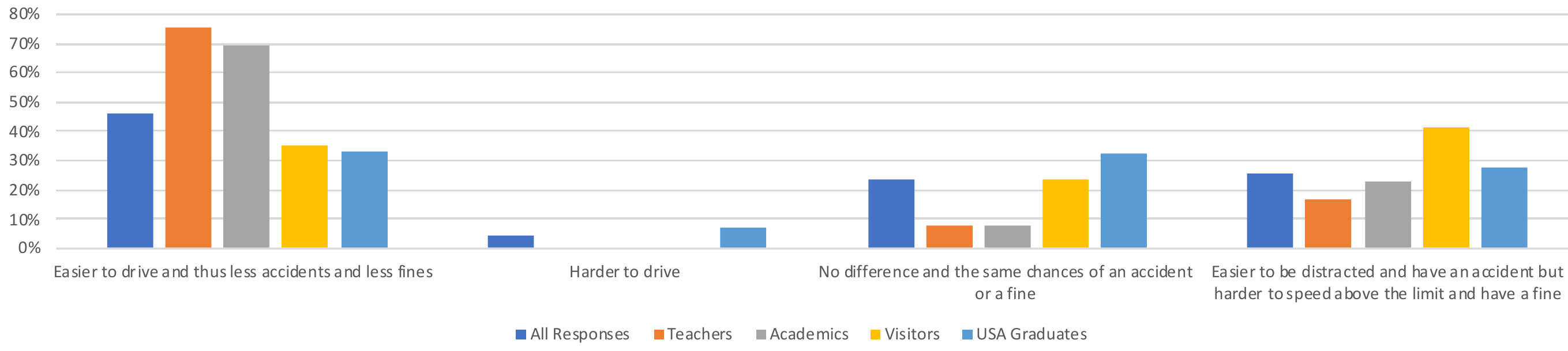}}
\caption{\vspace{-1mm}\label{Q10ANDQ15bars} Bar charts of responses
to Question~\ref{Q10} and Question~\ref{Q15}.
}
\vspace{-2.5mm}
\end{figure}
Figure~\ref{Question10} confirms that respondents concur on an  autopilot being
useful. The mode for all is that an autopilot is preferable (in Figure~\ref{Question10} we can see that all groups chose this option more than $50\%$ of the time except Open Day visitors, but even they have above $40\%$ for constant computer assistance). Almost no respondents would change the auto-pilot for a human co-pilot. Similarly, barely any respondent answered that a car is harder to drive with cruise control. Academics see no benefit from cruise control and a non trivial proportion of online respondents and Open Day visitors
see no impact of cruise control in reducing speeding fines or accidents (refer to Figure~\ref{Question15}).

While cruise control and autopilot originally involved mechanical devices, they are now operational through digital micro-controllers. We explored the issue of ``thinking machines''~\cite{Vega,Bibel2020} with what has been traditionally
perceived as an intellectual challenge (even after the birth of artificial
intelligence). Therefore, Question~\ref{Q17} was designed expecting the vast majority of respondents accepting that mechanized reasoning and performance by artificial chess-players is superior to human performance. Figure~\ref{Question17} shows that this is the case (especially academics), but even online respondents' responses show less than $4\%$ believe that chess programs are worse than the average human.
\begin{figure}
\centering
\subfloat[\label{Question18}Question~\ref{Q18}: Has ICT and search engines provide and advantage to compete in Trivial Pursuit.]{
\includegraphics[width=0.45\textwidth]{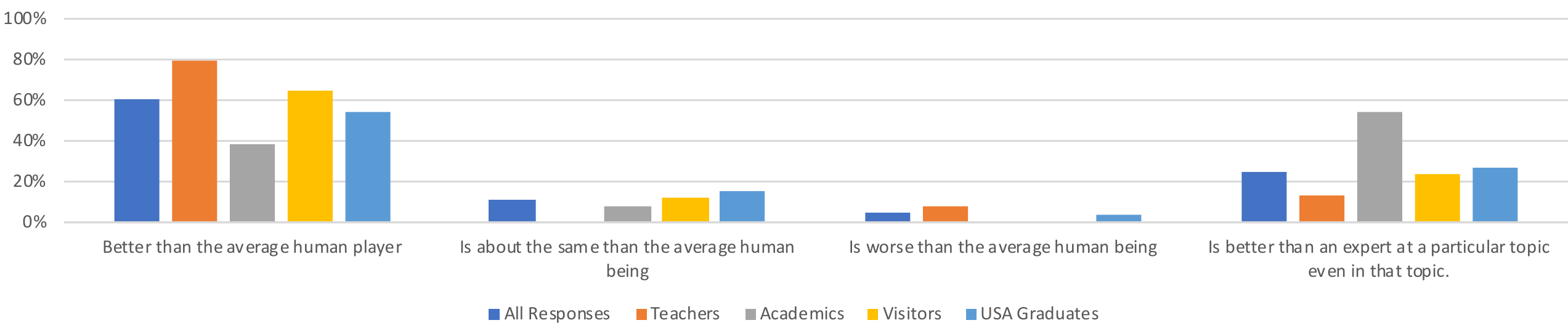}} $\;\;$
\subfloat[\label{Question19}Question~\ref{Q19}: Has ICT improved the performance of office workers.]{
\includegraphics[width=0.45\textwidth]{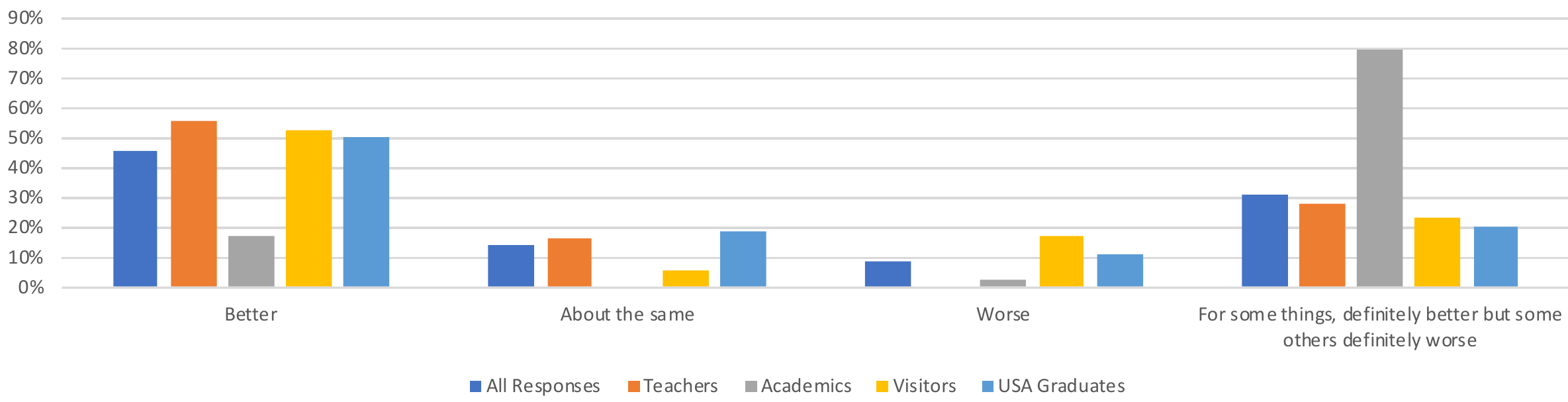}}
\caption{\vspace{-1mm}\label{Q18ANDQ19bars} Bar charts of responses to Question~\ref{Q18} and Question~\ref{Q19}.
}
\vspace{-2.5mm}
\end{figure}

\subsubsection{Analysis as to whether justifications are necessary}
\begin{figure}
\centering
\subfloat[\label{Question22}Question~\ref{Q22}: Argument for no transparency on intellectual property grounds.]{
\includegraphics[width=0.45\textwidth]{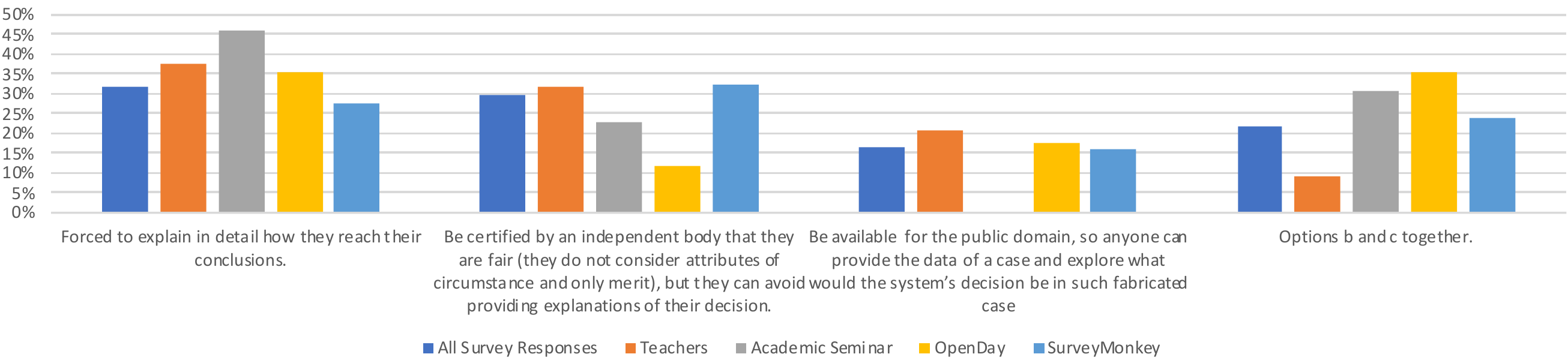}} $\;\;$
\subfloat[\label{Question23}Question~\ref{Q23}: AI-judge seems fairer than human judges.]{
\includegraphics[width=0.45\textwidth]{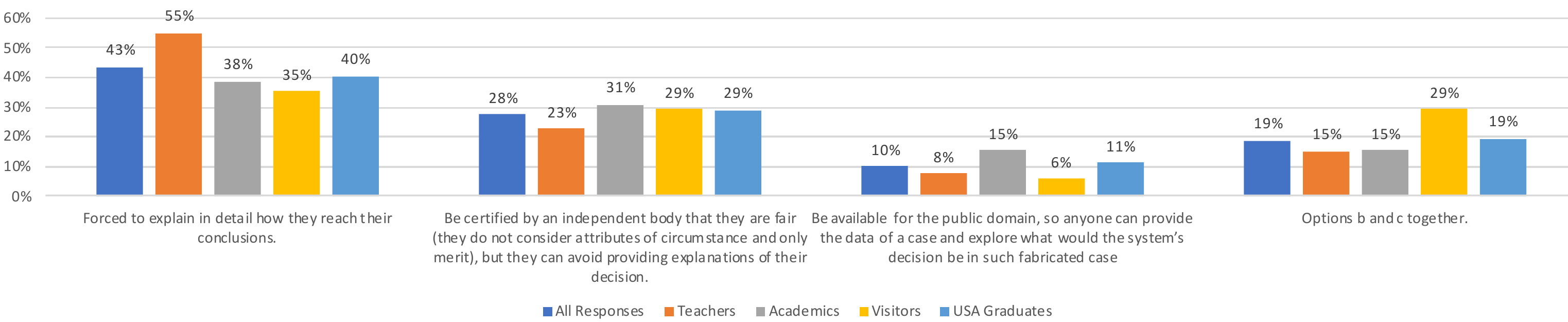}}
\caption{\vspace{-1mm}\label{Q22ANDQ23bars} Bar charts of responses about whether  AI-judges should  offer explanations and justifications for their decisions (Question~\ref{Q22} and Question~\ref{Q23}).
}
\vspace{-2.5mm}
\end{figure}

Since the deployment of expert systems,  it became clear that explanations~\cite{Darlington2013AspectsOI} are crucial for human users' 
trust~\cite{Swartout1993,WangPH2016} and acceptability of these AI systems~\cite{YeJ95}. We formulated two questions regarding the transparency of the decision provided by artificial agents. The European Union considers transparency---including traceability, explainability, and communication---as central issues for Trustworthy AI~\cite{EUReport}. In these questions (Question~\ref{Q22} and Question~\ref{Q23}) the categories are not ordinal. We offered one option for artificial agents being forced to explain, communicate, and justify their decisions. The other three options were such that the AI-Judge would not be required to provide an explanation. One was simply a certification that the system is fair by some third party, while the second was that the system could be used publicly as a black box. The third option for the non-transparent version was to combine the two previous ones. An interesting result is that online respondents do not seem to care about justification for a decision. Responses for forcing the system to explain and justify itself was less than $28\%$. Although the mode for teachers and academics is above 40\%, demanding AI decision-makers to provide explanations, these participants (in some earlier questions) seemed more accepting of an alien/artificial mind.
Teachers and academic responses do not reach $50\%$ for demanding explanations from AI for Question~\ref{Q22} (see Figure~\ref{Question22}).
This low interest for transparency (as per  Question~\ref{Q22}) are in contrast with those of  Question~\ref{Q23} where the requirement for an AI-judge to
explain and justify itself almost reaches $50\%$ for all collection approaches
(see Figure~\ref{Question23}). The mode for all participant groups is that artificial agents must be forced to justify and explain their decision making.

\textbf{Discussion:}
 Perhaps Question~\ref{Q23}  is more specific, as it refers to a particular individual and the particular situation of ruling on the lengths of a sentence for a convicted criminal. It is common that humans offer somewhat contradictory views for seemingly similar situations, as individuals value the specifics of each scenario differently.

\subsubsection{Analysis as to whether randomization is acceptable}
\begin{figure}
\centering
\subfloat[\label{Question13}Question~\ref{Q13}: If selecting which pedestrian to harm, what provides better randomness.]{
\includegraphics[width=0.45\textwidth]{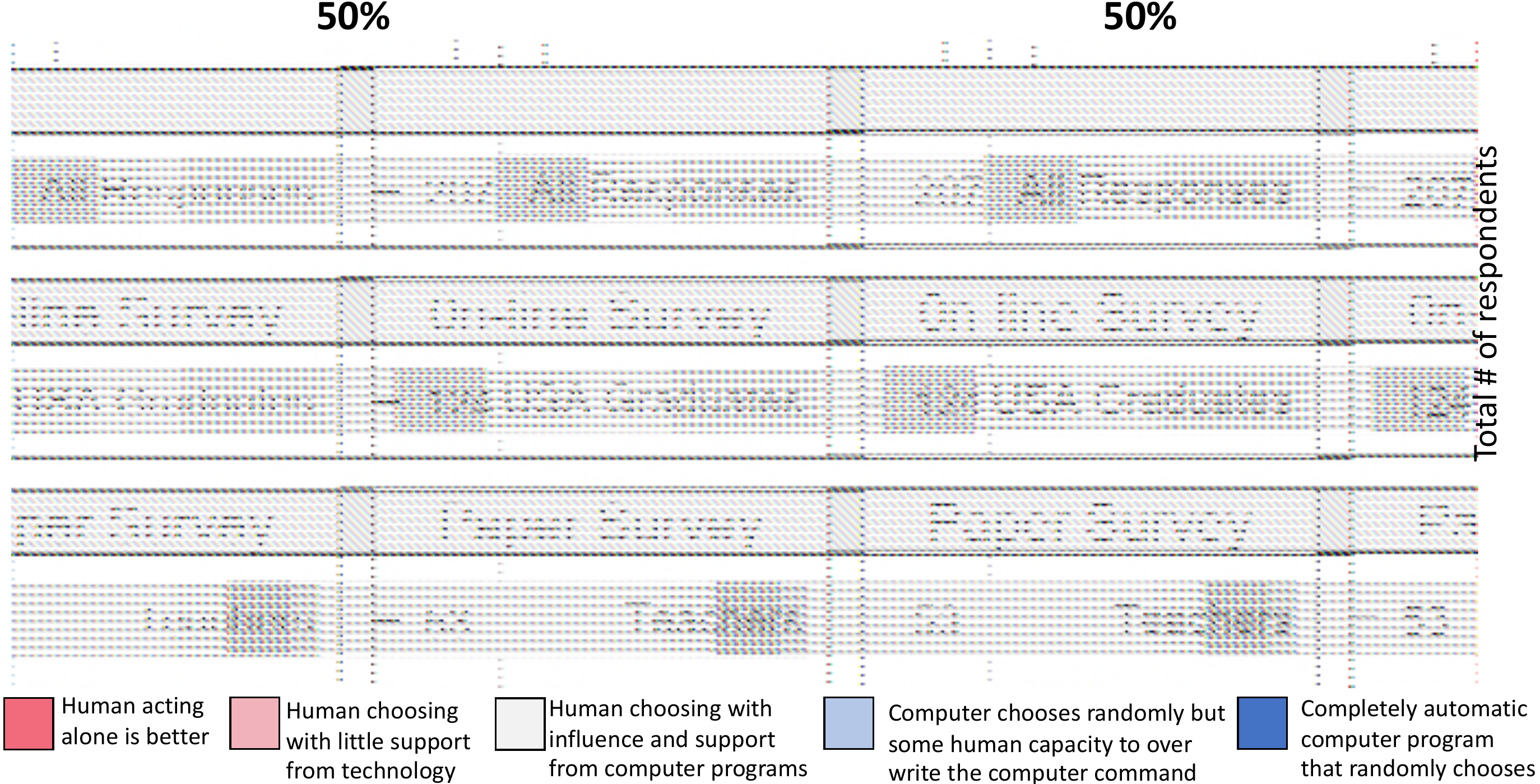}} $\;\;$
\subfloat[\label{Question20}Question~\ref{Q20}: If selecting which passenger to eject from a flight, what provides better randomness.]{
\includegraphics[width=0.45\textwidth]{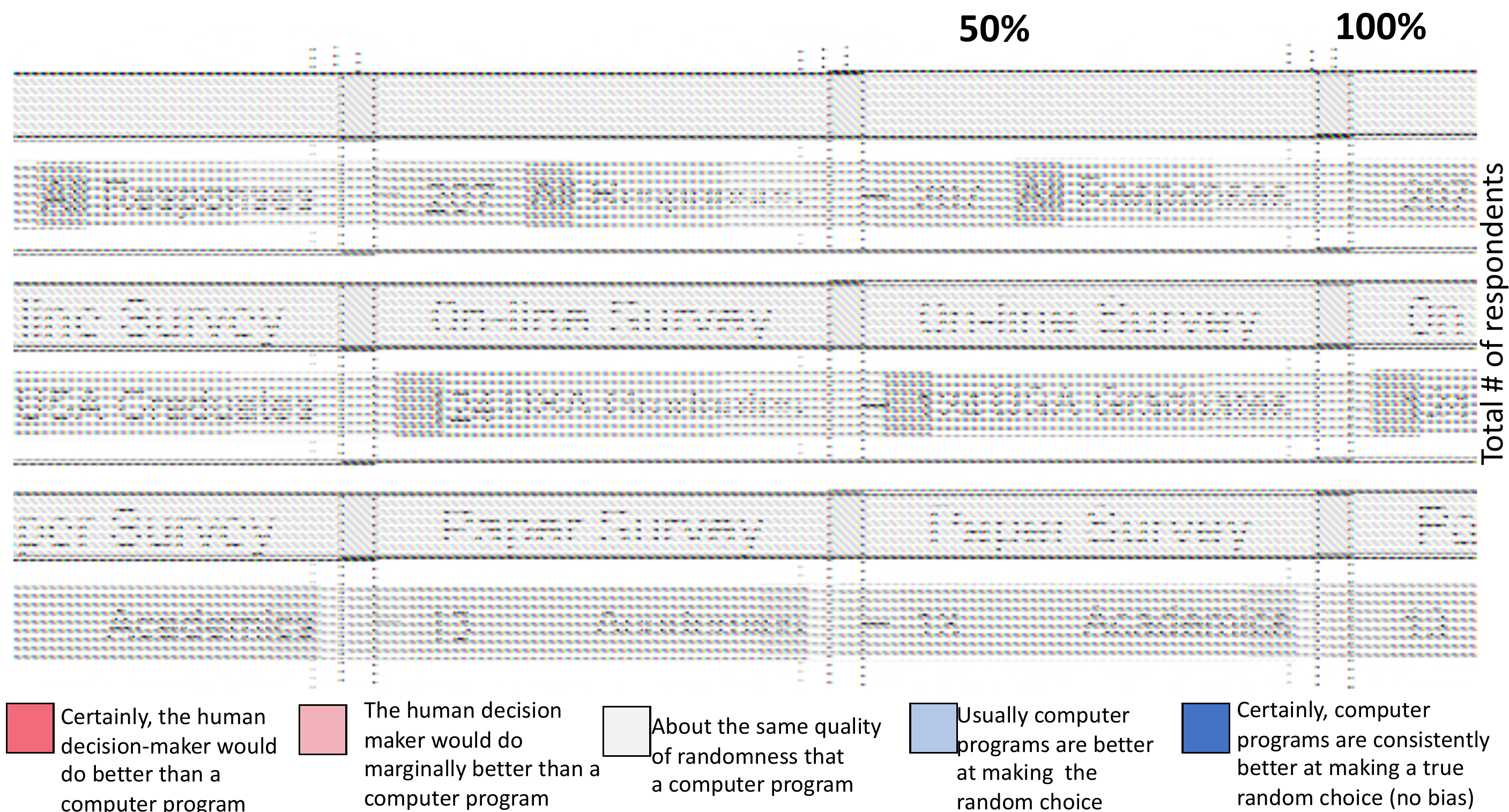}}
\caption{\vspace{-1mm}\label{Q13ANDQ20bars} Divergent bar charts of responses
to Question~\ref{Q13} and Question~\ref{Q20}.
}
\vspace{-2.5mm}
\end{figure}

For Question~\ref{Q13} and Question~\ref{Q20}, we were expecting responses to favor automatic choice since the questions suggest that a sensible choice is a random choice. Another reason for this expectation was that our earlier questions (Question~\ref{Q1} and Question~\ref{Q3}) show that our participants do regard computers as effective in calculations. As we argued earlier, if the merit of the case for two humans cannot be determined, a fair choice is a random choice. These two questions have been worded so as to suggest that computers
would be of valuable assistance even when making random choices. We find 
responses to Question~\ref{Q13} (see Figure~\ref{Question13}) are not inclined to a particular level of machine intervention. The results indicate a preference for computer involvement in such randomized decision-making alongside significant human participation. Only for the teachers,  the median is to use a computer's random choice and this is statistically significant over online respondents responses ($U=2,040$ and $p=0.0003$).
Question~\ref{Q20} recalls the scenario of Question~\ref{Q3}, where airline
staff ejects some passenger from a plane. The scenario now  insinuates that a random choice is required, and the responses here favor a strong use of computers in sharp contrast to the results we just described for Question~\ref{Q13}. In this setting, only less than $13\%$ of the online respondents chose the human alone option. Teachers, Open Day visitors, and academics are decisively for computer use. The option where the human decision-maker would do marginally better than a computer program received no responses from the teachers, no response from the academics, and only one from the Open Day visitors. Comparing teachers' responses to Question~\ref{Q20} with those to Question~\ref{Q13}, we find they are statistically significant ($U=923$ and $p=0.0118$). Similarly, academics' responses to Question~\ref{Q20} and  Question~\ref{Q13} ($U=38$ and $p=0.0914$) and Open Day visitors' responses to Question~\ref{Q20} and  Question~\ref{Q13} ($U=78$ and $p=0.116$) are also statistically significant.

\begin{figure}
\centering
\includegraphics[width=0.45\textwidth]{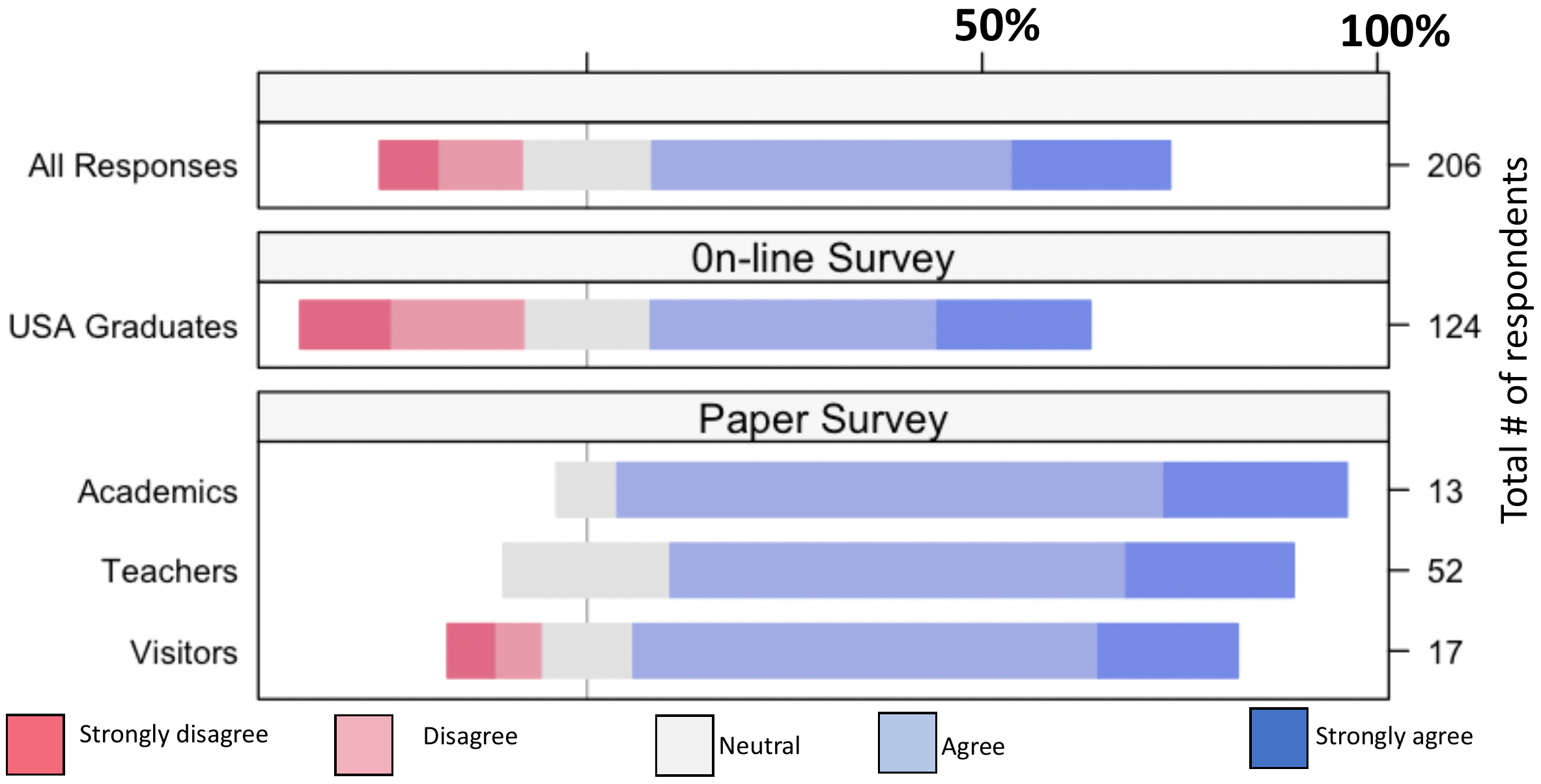}
\caption{\vspace{-1mm}\label{Question24divergent} Responses
to Question~\ref{Q24}: Computers would be able to generate random numbers to make fair decisions when the merit between two people is equal.
}
\vspace{-2.5mm}
\end{figure}

\textbf{Discussion:}
Note that Question~\ref{Q13} concerns a scenario where a decision results in a human being suffering some physical  harm. Under such circumstances, exclusively human and fully automated decisions are the least preferred choices. However, the three groups (academics, teachers, and Open Day visitors) display an absolute and radical shift in favor of computer/machine randomness when the issue (Question~\ref{Q20})  does not seem to be a matter of life or death. It seems that human preferences for AI decisions are dependent on the issues at stake. Question~\ref{Q24} offers a Likert scale with the standard options  (Strongly Disagree, Disagree, Neutral, Agree, Strongly Agree) regarding the respondents'  belief that computers can obtain fair random numbers for an unbiased choice when merit is equal.  Figure~\ref{Question24divergent} shows respondents strongly believe that computers can reproduce a truly random generator.

Although online respondents have a median and a mode that agrees on computers ability to achieve randomness, from a Mann-Whitney test we get a $p$-value of $0.2183$. Thus, we can reject the null hypothesis that the online respondents and the Open Day visitors have the same scoring tendency at the $5\%$ level.  Teachers and academics clearly believe even stronger (see Figure~\ref{Question24divergent})  in  achieving fair decisions with random generators in computers.  It is not surprising that respondents differ when discussing randomness, seeing as there are several accepted mathematical approaches on the issue~\cite{Volchan}.

Finally, Question~\ref{Q25} and Question~\ref{Q26} offer only yes/no options. Question~\ref{Q25} aimed to find out if participants are aware of CPUs' capacity to use thermal noise within the silicon as an entropy source. Overwhelmingly, our respondents answered no. Only two of the academics were aware of this issue. Question~\ref{Q26} tested for participants' awareness of the controversy around replacing the random oracle model with cryptographic hash functions.  Again, positive responses were minimal (below $1\%$) expect for the academics, where $2$ in $13$ (i.e., $15\%$) indicated their knowledge of the random oracle controversy.

\textbf{Discussion:}
Overall, it would seem that the public has trust in machines exhibiting non-deterministic and random behavior with what seems like little understanding of implementation issues.  Indeed, people are generally not particularly bothered with how machines work as long as their use is socially considered an improvement:
\begin{quote}
\emph{``Risks do not militate against the introduction of automated driving if the balance of risks is fundamentally positive''}~\cite{GERMANreport2017}.
\end{quote}

\section{Conclusions}
\label{Conclusion}

This paper has argued that if AI systems are to be involved in the automation of decisions affecting human lives, such decisions would need to be randomized for a significant number of scenarios. We also noted that this would revolutionize the notion of machine (automata): Determinism and deductive reasoning would alternate with random choices of carefully learned distributions. AI systems would give the impression of possessing some sort of free spirit, despite consistently following the law and making information-based decisions. Importantly, their decisions would even be potentially explainable and rationalizable (in the sense of game theory) and justifiable (in the sense of human interpretability). Potentially, most people could grow truly comfortable with AI decisions, perceiving them as fairer and less biased than human decision-makers. Others would likely find the idea of giving chance such a central role in decisions that so crucially affect their lives  highly unsatisfactory.

We have mathematically proven the validity of our arguments for random classifiers in the supervised learning setting, and also furnished game theoretic evidence for the superiority of mixed strategies under partial information assumptions. We have also provided empirical evidence to support that there is at least some societal acceptance for deploying random classifiers. Our results show that while people tend to favor partially or even fully automated decisions over human decisions in many circumstances, 
there is reluctance to resort to randomized decisions in life-or-death situations. Finally, we also discussed some policy and implementation issues with a particular focus on potential tradeoffs in connection with ever-increasing transparency requirements and current challenges in generating truly random sequences, to contribute to relevant debates in current AI policy and standardization initiatives.

Our aim was to provide insights that are useful both for AI researchers and policymakers. Researchers of various disciplines active in the AI field can benefit from our interdisciplinary approach, which links formal ML and game theory propositions with the state of art in AI ethics research and key current regulatory/policy considerations. Braking down traditional silos this way helps illustrate how insights from these disparate fields fit together in solving the important real-life problems we address.
 
On the policy side, we connect to key international AI policy and standardization initiatives. These initiatives increasingly attempt to guarantee the highest possible levels of transparency and explainability of AI systems to ensure accountability. While this is a commendable goal, our key message to policymakers is that maximizing transparency levels is neither the only nor necessarily the best solution. Instead, we need to keep potential drawbacks of transparency in mind and explore alternative accountability mechanisms that allow us to garner the benefits of randomization.

We believe the discussion here also raises a series of other interesting questions. For instance, do our findings also hold for other than supervised learning tasks?  On the game theory side, we indicated that AI interactions would be repeated games from a game-theoretic perspective and that human agents would be inclined to game the system if transparency requirements revealed the deterministic nature of the AI. Human agents would adapt. Should the AI adapt as well? Should it include ML modules for continuous learning? While the examples could go on, we hope to have sparked some interest in both the scientific and policymaking community to engage with these and many more related open problems.

{\small 
\bibliography{main}
\bibliographystyle{alpha}
}

\appendix
\section{Research Instruments}
\label{ap_research_inst}

\subsection{Introduction to Survey Respondents}
\label{EOPintro}

{\small
\underline{INSTRUCTIONS}: Please review the information sheet with details of the aims of the research,  the contact data of the chief investigator, and the task.  Please note that participation is voluntary, you can stop at any time, and your answers will  remain completely anonymous. 

Below are a series of questions that present scenarios where a professional usually makes  a decision that may impact on another person. You will be asked to grade the suitability of the decision being performed by the  decision-maker completely unassisted on one extreme to a computer program with no human intervention  in the other extreme. You should answer what suits best your current moral principles and values.  The time it takes you to complete the survey is irrelevant.

Equality of opportunity (EOP) is a widely supported ideal of fairness;  an equal opportunity policy seeks to create a level playing field among individuals,  after which they are free to compete for different positions. The positions that individuals earn under the condition of equality of opportunity reflect their  merit or deservingness, and for that reason, inequality in outcomes is considered ethically acceptable.

An individual's outcome/position is assumed to be affected by two main factors: 
his/her circumstance $C$ and merit $M$.  Circumstance $C$ is meant to capture all factors that should deemed irrelevant,  or for which the individual should not be held morally accountable.  The individual can hardly change those factors.  For instance, $C$ could specify the socio-economic status he/she is born into,  their gender, the colour of their skin. 

On the other hand, merit $M$ captures effort, discipline, and all those attributes that highlight  the good character of the individual.  Merit $M$ captures all accountability factors --- those that can morally justify inequality. 

For instance, within the context of employment,  one (narrow) interpretation of EOP requires that desirable jobs  are given to those persons most likely to perform well in them, because they have worked hard  at becoming experts, skilful and knowledgeable in the tasks required  for the job (and they shall not get the job because they are related to a stakeholder in the employment agency).  This interpretation of EOP allows for native talent and ambition to justify inequality in social positions,  whereas circumstances of birth and upbringing such as gender, race, and social background cannot. 

For any circumstance $C$ and any merit level $M$, a policy induces a distribution of utility among  people of circumstance $C$ and merit $M$. 
Formally, an EOP policy will ensure that an individual’s final utility will be, 
to the extent possible, only a function of their merit and not their circumstances. 
}

\subsection{Questionnaire}
{\small
\begin{enumerate}
\item 
\label{Q1}
Do you believe human beings are better (humans are fairer) than computers at applying an Equality of Opportunity (EOP) policy?
\begin{enumerate} 
\item Humans are never fairer, not at all   
\item Humans are slightly worse at being fair    
\item 	Neutral   
\item Humans are slightly better at being fairer, computers have sometimes a bias   
\item Definitely, humans are always better than any automation of an EOP policy.
\end{enumerate}
\item 
\label{Q2}
	Suppose you are learning a foreign language (for example French),  and your level on that language is being assessed by a human teacher of that language.  How fair you think that persons would judge your ability in the foreign language (your merit alone and not factors like your cultural background or your age,  which you cannot change) and how much would you prefer to be assessed by a computer program?
\begin{enumerate} 
\item 	Human judgement alone   
\item  	Human judgement with little support from computer programs
\item   Human judgement with support from computer programs        
\item   A computer program with some human monitoring   
\item  	Completely automatic computer program that uses only your merit as input.
\end{enumerate}
\item 
\label{Q3}
Suppose that you are a passenger in an air-plane and there is not enough fuel,  so the airline is forced to remove some travellers and offer them a place in a flight the next day.  How fair do you think airline officials would be in choosing the people who are  to stay behind in using criteria based on Merit and not on factors they should ignore, such as age or race?\\
(same options as Question~\ref{Q2})
\item 
\label{Q4}
Suppose because of your work and your spouse's work you are forced to enrol your child at  a school where the child is a visible minority (either because of religion or because of cultural background). How fair do you think the principal or the teachers would be when there is a conflict between your child  and other students which deserves some disciplinary action?  What would inspire you the most confidence regarding the fairness of the punishment to your child?
\begin{enumerate} 
\item   Human judgement alone   
\item   Human judgement with little support from 
\item   Human judgement with support from computer programs               
\item   A computer program with some human monitoring   
\item   Completely automatic computer program that uses only the child's merit as input.
\end{enumerate}
\item 
\label{Q5}
Suppose you are stranded in a shopping centre late at night in an area where you would be  noticeably identified as a minority (because of race, skin colour, religion or cultural heritage),  you need to plea for help to restart your car or to borrow a phone to make a call. In which case would you receive a fair go of help, when the security officer decides whether  you are a genuine person in need of help or the security officer uses technology to identify  your need of help (and the software is programmed not to consider circumstances such as race or cultural background)?\\
(same options as Question~\ref{Q2})

\item 
\label{Q6}
Suppose you are a patron in a night club in another country. You are having a great time, and you are not abusing alcohol or using recreational drugs. You are playing by the rules, but you are a visible minority because you are a tourist.  How much would you trust the bouncers (employees in charge of ensuring adequate behaviour) to fairly judge your behaviour and not expel you from the premises?  Or would you trust a software tool as fairer to judge whether your behaviour is adequate  when you know the software does not use any attributes about
 cultural background or issues such as race, religion, etc.?\\
(same options as Question~\ref{Q2})
\item 
\label{Q7}
Suppose you are to adopt a child (a baby, for example), you are being interviewed by family-services staff.  How much would you believe their decision to allow you to adopt would be fair and removed from circumstantial  factors, while actually based on merit for your potential as a parent if they acted with no support  from information technologies, or would the decision be fairer if it was a software that explicitly  focus on merit (and not circumstance such as race, or socio-economic background)\\
(same options as Question~\ref{Q2})
\item 
\label{Q8}
Consider a judge in a hearing for bail of an accused of a violent crime. How fair would the judge be in making the decision to award bail on the merits of the case  and without considering circumstantial issues such as race, gender, socio-economic background? Grade your answer relative to a software tool that has as input only the current evidence on the case?
\begin{enumerate}
\item Human judgement alone is the fairest   
\item Human judgement with little support from technology is the fairest   
\item Human judgement with support from computer programs is the fairest    
\item  A computer program with some human supervision is better   
\item A completely automatic computer program that uses only the merit of the case is the fairest 
\end{enumerate}
\item 
\label{Q9}
Consider you have some discomfort and have suffered from the same chronic illness from some time. Describing the symptoms to a pharmacist is difficult because of the potential judgement of your personal and private life. Would you prefer to request the corresponding medication to a robot or software who does not have a body and cannot understand the issues? Would you trust that the pharmacist would not make some judgement on your personal life because of the type of medication you are requesting?
\begin{enumerate} 
\item I would feel completely no embarrassment in discussing the medication with a human pharmacist face–to-face
\item I would prefer some technology support so I would not need to discuss or request the medication with a human pharmacist face–to-face
\item I believe human judgement will happen even with support from computer programs       
\item I would be more comfortable providing input to a computer program with some human supervision so customers do not abuse prescriptions   
\item I would prefer a completely automatic computer program that I can even order online when the medication does not require prescription and in any case only the prescribing doctor needs to know my personal situation.
\end{enumerate} 
\item 
\label{Q10}
Consider the scenario of a pilot in the control of a large passenger plane on a trans-oceanic flight. In your opinion, the pilot performs better with
\begin{enumerate} 
\item The constant assistance of computers for guidance but no help with another person    
\item A co-pilot and no assistance from computers to any of them     
\item Computer system that can fly the plane in autopilot and enable the pilot to take a rest       
\item Only some instruments that measure height, velocity and heading but no autopilot.
\end{enumerate} 
\item 
\label{Q11}
Suppose you are applying for a visa to travel to another country.  Would you prefer your application to be assessed by a panel of people or would you  prefer that the awarding of the visa be decided by some software that clearly  indicates what are the inputs and does not consider confidential information?  Do you think the decision would be better based on merit according to the level of automation?\\
(same options as Question~\ref{Q2})
\item 
\label{Q12}
Today, credit card applications are evaluated by software programs.  Rarely a human banker is involved in deciding whether the applicant should be awarded a credit card and  even what credit limit.  Do you think credit card applications were awarded more fairly in the past when the applicant was  interviewed by a bank-credit officer or do you believe they are now more fairly assessed on merit alone (and not circumstantial factors of discrimination like race, or even the clothing or personal presentation that may suggest a specific cultural background)?
\begin{enumerate} 
\item Human judgement alone   
\item Human judgement with little support from technology    
\item Human judgement with support from computer programs       
\item Computer program with some human monitoring the overall credit lender position but no need to check each individual case.   
\item Completely automatic computer program that uses only your merit as input
\end{enumerate} 
\item 
\label{Q13}
Suppose a software is controlling a self-driving car (also known as an autonomous vehicle),  and unfortunately ahead of the two pedestrians are standing on the road. One pedestrian on each of the two available lines.  The software of the car has a split of a second to decide which  line to drive basically resulting in two exclusive actions,  each would cause harm to one of the pedestrians, but save the other.  Because the car cannot evaluate the merit (or blame) of why each pedestrian is there, it makes a random decision given equal chance to each pedestrian to not be injured.  Do you think a human driver would be able to make the same or even better random choice?
\begin{enumerate} 
\item Human acting alone is better
\item Human choosing with little support from technology    
\item Human choosing with influence and support from computer programs       
\item Computer program chooses randomly but some human capacity to over-write the computer command. 
\item Completely automatic computer program that randomly chooses.
\end{enumerate} 
\item 
\label{Q14}
Suppose the parents’ committee of a school is applying for a government grant to carry out some  improvements to the classrooms and the learning materials of the school.  How likely do you believe is the panel of humans to review the application fairly and based on merit  regarding the schools learning outcomes and not biased from circumstance issues such as the cultural  profile of families' cultural and religious background whose children attend the school.\\
(same options as Question~\ref{Q2})
\item 
\label{Q15}
	A car with cruise control is
\begin{enumerate}
\item Easier to drive and thus less accidents and less fines  
\item Harder to drive 
\item No difference and the same chances of an accident or a fine       
\item Easier to be distracted and have an accident but harder to speed above the limit and have a fine
\end{enumerate}
\item 
\label{Q16}
	Self-driving cars (autonomous vehicles) would be safer than
\begin{enumerate}
\item Cars driven by experienced human drivers
\item Cars driven by novice human drivers
\item Professional drivers such as taxi drivers
\item Human drivers with several categories of driver licenses
\item None of the above
\end{enumerate}
\item 
\label{Q17}
An engineered (with artificial intelligence) software's performance at a game such as chess is
\begin{enumerate}
\item
	Better than the average human player
\item Worse than the average human player
\item About the same as the average human player
\item Typically, better than an expert human player (a grand master)
\end{enumerate}
\item 
\label{Q18}
Search engines' (such as GOOGLE) performance for finding (sort of remembering) 
facts (for example, consider someone playing Trivia with the support or assistance of the World-Wide-Web i.e.  the Internet and Google versus participants at Trivia without such help)
\begin{enumerate}
\item
Is better than the average human being
\item
Is about the same than the average human being
\item
Is worse than the average human being
\item
Is better than an expert at a particular topic even in that topic.
\end{enumerate}
\item 
\label{Q19}
With the assistance of digital calendars (with features such as reminders and addresses of contacts),  an office workers perform their job relative to no digital tools
\begin{enumerate}
\item
Better
\item
About the same
\item
Worse
\item
For some things, definitely better but some others definitely worse
\end{enumerate}
\item 
\label{Q20}
Suppose again that you are a passenger in an air-plane and there is not enough fuel,  so the airline is forced to remove some travellers and offer them a place in a flight the next day.  Suppose several passengers are assessed to have equal low Merit.  Do you think a human decision-maker would do a random choice (equal probability) among equal-merit passengers?
\begin{enumerate}
\item 
Certainly, the human decision-maker would do better than a computer program
\item 
The human decision-maker would do marginally better than a computer program    
\item 
About the same quality of randomness that a computer program   
\item 
Usually computer programs are better at making the random choice   
\item 
Certainly, computer programs are consistently better at making a true random choice (no bias)
\end{enumerate}
\item 
\label{Q21}
In 2017, a man named Eric Loomis was sentenced to six years in prison as this was the sentence  recommended by an AI algorithms and the preceding judge who confirm it. Do you think the sentence to a person found guilty (besides meeting the parameters of the law)  should be performed by
\begin{enumerate}
\item
A human judge with no assistance from technology
\item
A human judge with some assistance from technology
\item
A human judge in conjunction with assistance from technology
\item
A computer program that uses artificial intelligence with some monitoring and supervision by a judge. 
\item
A computer program that uses artificial intelligence and no supervision by a judge.
\end{enumerate}
\item 
\label{Q22}
The industry of machine learning and artificial intelligence argues that their smart artificial  decision-making software is confidential intellectual property. Should computer algorithms that make decision affecting human beings be
\begin{enumerate}
\item
Forced to explain in detail how they reach their conclusions.
\item
Be certified by an independent body that they are fair (they do not consider attributes of circumstance and only merit), but they can avoid providing explanations of their decision.
\item
Be available for the public domain, so anyone can provide the data of a case and explore what would the system’s decision be in such fabricated case
\item
Options (b) and (c) together.
\end{enumerate}
\item 
\label{Q23}
In Broward County, Florida, USA more than 7,000 defendants faced a bail hearing where a  software predicted the likelihood to commit a crime again if the defendant was released.  Based on the software's recommendation, the presiding judge made a recommendation. A study by Pro-Publica found that the algorithm refused bail to black defendants far more  often that human judges, but the workings of the algorithm are kept secret by their developers.  Should computer algorithms that make decision affecting human beings be\\
(Same options as Question~\ref{Q22})
\item
\label{Q24}
Computers would be able to generate random numbers to make fair decisions when the merit between  two people is equal.
\begin{enumerate}
\item Strongly disagree
\item Disagree
\item Neutral
\item Agree
\item Strongly agree
\end{enumerate}
\item
\label{Q25}
Are you aware of the CPUs capacity to use thermal noise within the silicon as an entropy source and what does this imply, for example for reseeding a pseudo-random generator?
\begin{enumerate}
\item No
\item Yes
\end{enumerate}
\item
\label{Q26}
Are you aware of the controversy regarding the random oracle model  use for formal proofs of cryptographic protocols and its replacement in practices by cryptographic hash functions?
\begin{enumerate}
\item No
\item Yes
\end{enumerate}
\item 
\label{Q27}
Please indicate your highest completed education level
\begin{enumerate}
\item
Primary school
\item
Secondary school or high school
\item
Trades or technical qualification
\item
University degree
\item
Postgraduate degree (master, doctorate)
\end{enumerate}
\item 
\label{Q28}
	Please indicate your age range
\begin{enumerate}
\item 18-29
\item
30-39
\item
40-49
\item
50-59
\item
60 or above
\end{enumerate}
\item 
\label{Q29}
Are you currently
\begin{enumerate}
\item
Employed full-time
\item
Employed part-time
\item
Employed on casual work
\item
Self-employed or business owner
Retired
\item
Unemployed
\end{enumerate}
\item 
\label{Q30}
Would you consider yourself currently a minority in your work environment?
\begin{enumerate}
\item Yes
\item No
\end{enumerate}
\item 
\label{Q31}
Would you consider yourself currently a minority in your social environment?
\begin{enumerate}
\item Yes
\item No
\end{enumerate}
\item 
\label{Q32}
How do you engage in policy-making with your government? (In this question, you can chose more than one option)
\begin{enumerate}
\item
I do not vote
\item
I do vote
\item
I am a member of a political party
\item
I correspond with my member of parliament (regularly, seldom, few occasions, never)
\item
I respond to invitations for comment
\item
I am a member of a union or other body that is not a political party but engages to influence policy-making
\item
I regularly run a blog/radio show/magazine editor and consider myself an influencer in social media/paper/media/broadcasting that may influence public opinion and/or policy-making
\item
I directly work for government in advice, consulting or the like for policy-making
\item
I participate in rallies, demonstrations, and other event with massive attendance aimed at influencing policy-making
\item
I sing petitions, and similar documents that aim to influence policy-making.
\item
I respond to government's official inquiries or I participate in their preparing responses
\item
Any other you are actively involved \hrulefill
\end{enumerate}
\end{enumerate} 
}

\section{Declarations}
The authors have not received any funds, grants, or other support, and have no relevant financial or non-financial interests to disclose.
This study was performed in line with the principles of the Declaration of Helsinki. Approval was granted by the Ethics Committee of
Griffith University GU~Ref~No:~2019/617.
Informed consent was obtained from all individual participants included in the empirical study.
The anonymized responses of the empirical study can be made available  in a public repository.

\end{document}